\documentclass[prc,href,aps,a4paper,groupedaddress,nofootinbib,twocolumn]{revtex4}
 \usepackage{hyperref}
 \usepackage{graphicx}
 \usepackage{amsfonts}
 \usepackage{amssymb}
 \usepackage{amsmath}
 \usepackage{natbib}
 \usepackage{dcolumn}
 \usepackage{bm} 
 \usepackage{multirow}
 \usepackage{epsfig,ulem}
 \usepackage{amsfonts}
 \usepackage{hhline,color}
 \usepackage{pifont}
 \usepackage{lmodern}
 \usepackage[utf8]{inputenc}

\begin{document}
\title{ Estimating nuclear equation of state parameters away from saturation density}

\author{Naosad Alam}
\email{naosad.alam@tifr.res.in}
\author{Subrata Pal}
\email{spal@tifr.res.in}
\affiliation{Department of Nuclear and Atomic Physics, Tata Institute of Fundamental Research, Mumbai 400005, India}

%\date{\today}
\begin{abstract}
We explore the density variation of the correlation coefficient of the key parameters of the nuclear equation 
of state (EoS) with the bulk and crustal properties of neutron stars. The analysis was performed using two
diverse sets of nuclear effective interaction theories based on nonrelativistic Skyrme-Hartree Fock model 
and relativistic mean field model. We find that the commonly studied EoS parameters, namely the isoscalar 
incompressibility of symmetric nuclear matter $K(\rho)$ and the isovector slope of symmetry energy $L(\rho)$, 
reveal consistently maximum correlation with the radius, tidal deformability, and moment of inertia all 
around twice the saturation density. We find even more tighter and robust correlations beyond the saturation 
density for constructed parameter $\eta = [KL^2]^{1/3}$ allowing the possibility to impose stringent constraints
on high-density $K(\rho)$ and $L(\rho)$. Extensive correlation analysis of the EoS parameters with the radius 
and tidal deformability bounds from the gravitational wave events and recent pulsar observations allow us to provide reliable constraints
on the central values of $K(\rho_0) \approx 240$ MeV and $L(\rho_0) \approx 48$ MeV at saturation density 
and $K(1.6\rho_0) \approx 332^{+88}_{-50}$ MeV and $L(1.6\rho_0) \approx 122^{+26}_{-18}$ MeV at 1.6 times the saturation density. 
The crust-core transition density and the crustal fraction of moment of inertia 
are shown to correlate moderately with $L(\rho)$ and $\eta(\rho)$ near the subsaturation density.

\end{abstract}
\maketitle

\section{Introduction}

The nuclear equation of state (EoS) has been the cornerstone in nuclear physics and astrophysics enabling accurate description of 
nuclear multifragmentation at subsaturation densities, the neutron skin and dynamics of heavy-ion collisions around the saturation 
density, and the properties of isolated neutron stars (NS) and composites formed in the merger of neutron stars and/or black holes 
at higher densities~\cite{MUSES:2023hyz,Sorensen:2023zkk,Stone:2021uju,Burgio:2021vgk,Baiotti:2019sew,Baldo:2016jhp,Watts:2016uzu,Lattimer:2006xb,Peilert:1989kr}. Inspite of considerable attempts to constrain the nuclear EoS by employing various combinations of nuclear 
experimental data, astrophysical observations, and theoretical modeling, the current knowledge of the EoS is largely ambiguous 
especially at supra-saturation densities~\cite{Stone:2021uju,Burgio:2021vgk}. 

In general, the EoS representing energy per nucleon $e(\rho, \delta)$ of neutron-proton asymmetric nuclear matter at a nucleon
density $\rho = \rho_n+\rho_p$ can be approximated by a parabolic expansion of isospin asymmetry $\delta=(\rho_n-\rho_p)/\rho$ as~\cite{Alam:2016cli,Vidana:2009is,Chen:2009wv}:
\begin{equation}
e(\rho,\delta) = e_0(\rho) + e_{\rm{sym}}(\rho)\delta^2  + {\cal O}(\delta^4).
\label{EOSANM}
\end{equation}
Traditionally, the neutron-proton symmetric part of the energy per nucleon $e_{0}(\rho)\equiv e(\rho,\delta=0)$ and the
nuclear symmetry energy $e_{\rm{sym}}(\rho)$ 
are parametrized/expanded about the nucleon saturation density $\rho_0$ via the dimensionless quantity
$\chi = (\rho -\rho_0)/3\rho_0$ to yield~\cite{Vidana:2009is,Chen:2009wv}
\begin{align}
\label{E0}
e_0(\rho) = & e_{0}(\rho_0)+ \frac{K}{2!}\chi^2 + \frac{Q}{3!}\chi^3 + {\cal O}(\chi^4), \\
e_{\rm{sym}}(\rho) = & e_{\rm{sym}}(\rho_0) + L\chi + \frac{K_{\rm sym}}{2!}\chi^2 + \frac{Q_{\rm sym}}{3!}\chi^4 
+ {\cal O}(\chi^6), 
\label{Esym}
\end{align}
While measurements of nuclear masses, density distributions and isoscalar giant monopole resonances of heavy nuclei 
led to reasonably accurate constraints on the $n-p$ symmetric EoS, namely the binding energy $e_0(\rho_0)$, 
incompressibility $K(\rho_0) = 9\rho_0^2[\partial^2 e_0(\rho)/\partial\rho^2]_{\rho_0}$ and 
skewness $Q(\rho_0) = 27\rho_0^3[\partial^3 e_0(\rho)/\partial\rho^3]_{\rho_0}$ only about the saturation density
$\rho_0 \approx 0.16$ fm$^{-3}$, the supranormal density information is scarce leading to diverse model predictions.

In contrast, inspite of intensive experimental and theoretical efforts, the nuclear symmetric energy $e_{\rm{sym}}(\rho)$
is largely uncertain even around the saturation density $\rho_0$ and depends on the methods and observables used in 
the estimation~\cite{Li:2021thg,Baldo:2016jhp}. Experimental measurements of the neutron skin thickness, hadron flow in heavy-ion collisions, 
isospin diffusion, isobaric analog states, giant and pygmy dipole resonances analysis are all found to be quite
sensitive to the symmetry energy and provided important constraints particularly on 
$e_{\rm{sym}}(\rho_0) = 30 \pm 4$ MeV. The deduced values for the slope
$L(\rho_0) \approx 30-87$ MeV, curvature $-400 < K_{\rm sym}(\rho_0) < 100$ MeV, 
and skewness $Q_{\rm sym}$ of symmetry energy 
however remain highly uncertain and poorly constrained even at $\rho_0$~\cite{Burgio:2021bzy,Tews:2016jhi, Zhang:2017ncy}.
On the other hand, recently quite precise and simultaneous measurements of mass and radius of neutron stars from Neutron Star Interior Composition
Explorer (NICER)~\cite{Riley2019,Miller:2019cac,Riley2021,Miller:2021qha}  and tidal 
deformability bounds from the detected gravitational waves GW170817~\cite{LIGOScientific:2017vwq,LIGOScientific:2018cki} and GW190814~\cite{LIGOScientific:2020zkf} 
from the merger of binary compact objects provide unique opportunities to explore the dense core of the compact objects.
Naturally, these astrophysical observables are influenced and expected to be strongly correlated with the EoS of 
asymmetric nuclear matter at higher densities beyond $\rho_0$.

The theoretical model analysis of EoS parameters are mostly confined at or below the saturation density. In particular,
extensive correlation studies/analysis has been carried out involving the key individual EoS parameters 
($K, L, K_{\rm sym}$ at $\rho \lesssim \rho_0$) and their specific combinations with the neutron star radii $R$, 
compactness $C \equiv M/R$, tidal deformability $\Lambda$, gravitational redshift, and crust-core transition densities $\rho_t$ \cite{Yang:2023uxq,Patra:2023jvv,Burgio:2021bzy,Carson:2018xri, Alam:2016cli,Malik:2018zcf,Fortin:2016hny,Ducoin:2011fy,Ducoin:2010as}.
All these studies which were performed either within a representative EoS or a large number of unified EoS, indicated 
moderate to strong correlations with the astrophysical observables thus allowing to put important constraints on the 
EoS parameters at densities $\rho \approx \rho_0$. To accurately constrain the EoS parameters at supra-saturation densities,
ideally one can perform the Taylor expansion of $e_0(\rho)$ and $e_{\rm{sym}}(\rho)$ about a $\rho > \rho_0$~\cite{Cai:2023pkt,Li:2021thg}. However, 
for the convergence of the series, one would require a multitude of parameters whose precise estimation is hindered 
by limited available data. Alternative approach to scan the density dependence of symmetry energy relies on finding
correlation between observables that are more sensitive to high density regime of the EoS ~\cite{Fattoyev:2014pja}.  Regardless of the analysis 
method or the models used, pronounced and robust correlations were found at a density $2\rho_0$ between the pressure $P$ 
of beta-equilibrated matter and radius of neutron star as $P \propto R^4$  \cite{Lattimer:2006xb,Lattimer:2000nx,Lim:2018bkq,Tews:2018iwm} as well as 
between the pressure and tidal deformability of a $1.4M\odot$ neutron star \cite{Tsang:2019vxn} and f-mode frequencies \cite{Kunjipurayil:2022zah}.

The symmetry energy $e_{\rm sym}$ and its slope $L$ also significantly influence the properties of neutron star crust 
at the subsaturation densities such as the composition, thickness, elasticity which in turn affect phenomena like 
cluster oscillations and gravitational wave emission. In fact, the crust-core transition density $\rho_t$
has been found to be well correlated with $L$ at $\rho \approx 0.10$ fm$^{-3}$~\cite{Ducoin:2011fy}. The crust-core transition density
determines the crustal pressure $P_t$ and moment of inertia, where the crustal fraction of the moment of inertia $\Delta I/I$ 
plays a crucial role in the study of pulsar glitches \cite{Link:1999ca,Espinoza:2011pq,Andersson:2012iu}. 

In this article, we have performed correlation analysis over a broad density range to explore crucial links between the
neutron star bulk, crustal properties (viz $M$, $R$, $\Lambda$, $\rho_t$) and the density variation of key EoS parameters, 
namely the incompressibility $K(\rho)=9\partial P/\partial \rho$ of symmetric nuclear matter, the symmetry energy slope $L(\rho)=3\rho \: \partial e_{\rm sym}/\partial \rho$ of asymmetric matter for proper characterization of the density dependence of the nuclear EoS. The incompressibility $K(\rho)$ determines the stiffness of symmetric nuclear matter, influencing the central density, overall compactness, and the maximum mass of neutron stars 
\cite{Alam:2016cli,Li:2021thg}. On the other hand, the slope parameter $L(\rho)$ governs the density dependence of the symmetry energy and is essential for modeling isospin-asymmetric systems, significantly affecting neutron star radii and proton fractions \cite{Burgio:2021bzy,Tews:2016jhi,Zhang:2017ncy}.
Together, these parameters regulate the pressure response at supranuclear densities for neutron-rich matter ($\delta \approx 1$) in the interior of neutron stars that determine several obervables such as the tidal deformability as well as the nuclear structure and reaction dynamics mainly
involving neutron-rich nuclei \cite{Sorensen:2023zkk,Baldo:2016jhp,Peilert:1989kr}.

To constrain the EoS parameters with a minimum uncertainty we pin down the density point/range which exhibit the strongest correlations. 
The primary objective is to extract information about the high-density region of the nuclear matter EOS by constraining its 
parameters beyond the saturation density based on their correlation with neutron star properties. 
Such a density variation of the correlation between the radii of neutron stars and the symmetry energy slope $L(\rho)$ has been 
analyzed using covariance analysis based on a single model \cite{Fattoyev:2014pja}. 
In the present calculations, we have employed a comprehensive set of relativistic mean field (RMF) theory~\cite{Dutra:2014qga} for nuclear interaction 
that provides Lorentz covariant extrapolation from sub- to supra-saturation densities as well as a representative set of 
nonrelativistic Skyrme-Hartree-Fock (SHF) model~\cite{Dutra:2012mb} that has a diverse high-density behavior and performed quantitative 
analysis using the Pearson correlation coefficient. 
Further, in search of a quantity that is strongly correlated with neutron star observables, we have focused on a new 
EoS parameter $\eta =[K L^2]^{1/3}$ that is constructed from a specific combination of nuclear EoS parameters \cite{Sotani:2013dga,Sotani:2015lya, Silva:2016myw}. The parameter $\eta$, defined as a combination of $K$ and $L$, effectively encapsulates their joint impact from both the isoscalar and the isovector part of the nuclear EoS on the properties of neutron stars. Constraining $\eta$ through the measurements of neutron star observables reduces the dimensionality of the nuclear matter parameter space, which further helps to constrain nuclear EoS models more efficiently.
At the saturation density, $\eta(\rho_0)$ was found to be strongly correlated with the mass, radius and the surface redshift of non-rotating neutron stars~\cite{Sotani:2013dga,Sotani:2020bey,Sotani:2022ucj}. Variations in $\eta(\rho_0)$ lead to a smooth change in the neutron star mass–radius relation~\cite{Sotani:2015lya}. It is therefore also of interest to explore whether such strong correlations in $\eta(\rho)$ do persist at higher densities, thereby allowing for a robust prediction/extraction of the individual EoS parameters $K(\rho)$ and $L(\rho)$ at $\rho > \rho_0$. 

The paper is organized as follows. In Sec. II, we discuss the EoSs employed in our correlation analysis.
Here, we briefly describe the formalism to calculate various bulk and crustal properties of neutron stars
with emphasis on the core-crust transition density. Sec. III contains our results and discussions on the correlations
between the EoS parameters and the neutron star observables. Finally, the conclusions are
drawn in Sec. V. We adopt the system of units $\hbar=c=G=1$ throughout the manuscript.

\section{EoS and properties of neutron star matter}

In this section we will discuss briefly on the nuclear models used and the bulk and crustal properties 
of the neutron star used in the analysis.

\subsection{Nuclear equations of state}
\label{sec:models}

To analyze the the properties of neutron star we have used  twenty-eight EoSs for beta-equilibrated star matter
based on the relativistic mean-field (RMF) \cite{Dutra:2014qga} and the non-relativistic Skyrme-Hartree (SHF) \cite{Dutra:2012mb} models.
In the original RMF model, the interactions between the nucleons are mediated by the exchange of 
scalar-isoscalar $\sigma$,  vector-isoscalar $\omega$ and vector-isovector $\rho$ mesons, 
with subsequent improvements via the inclusion of non-linear self- and cross-couplings between the 
mesons. The various RMF models used in the present calculations are the NL-type with nonlinear $\sigma$ 
interactions, NL3 \cite{Lalazissis:1996rd}, GM1 \cite{Glendenning:1991es};  NL3-type with additional $\sigma-\rho$ and $\omega-\rho$ terms 
NL3${\sigma\rho 4}$, NL3${\sigma\rho 6}$ \cite{Pais:2016xiu}, NL3${\omega\rho 02}$ \cite{Horowitz:2000xj}, 
NL3${\omega\rho 03}$ \cite{Carriere:2002bx}; TM-type with nonlinear $\omega$ terms 
TM1 \cite{Sugahara:1993wz}, TM1-2 \cite{Providencia:2012rx}; FSU-type with further nonlinear $\omega$ couplings
FSU2 \cite{Chen:2014sca}; and the BSR2, BSR3, BSR6 family of additional nonlinear couplings \cite{Dhiman:2007ck,Agrawal:2010wg}.

The SHF models used in this analysis are SKa, SKb \cite{Kohler:1976fgx}, SkI2, SkI3, SkI4, SkI5 \cite{Reinhard:1995zz}, Sly230a,  Sly230b \cite{Chabanat:1997qh},
Sly4, Sly5, Sly6, Sly7 \cite{Chabanat:1997un}, SkMP \cite{Bennour:1989zz}, KDE0V1 \cite{Agrawal:2005ix}, SK255, 
and SK272 \cite{Agrawal:2003xb}. 
The model coupling constants are obtained by fitting to finite nuclei and the infinite nuclear matter properties 
at $\rho_0$  and successfully described various experimental data for finite nuclei.
We have employed EoSs where the inner crust has been calculated assuming a polytropic form $P(\varepsilon)=a+b \varepsilon^{4/3}$ with $P$ and $\epsilon$ being the pressure and the energy density, respectively. The constants $a$ and $b$ are chosen so that, at one end, the EoS for the inner crust matches with the inner edge of the outer crust, and at the other end, it matches with the edge of the core.
The outer crust EoS is taken from the work of Baym-Pethick-Sutherland \cite{Baym:1971pw}.

\subsection{Structural properties of neutron stars}
\label{structure}  

For a given EoS, one can obtain the structural properties of neutron stars such as gravitational mass $M$
and the radius $R$ by solving numerically the following Tolman-Oppenheimer-Volkoff (TOV) equations 
\cite{Weinberg72,Glendenning97,Haensel07} which describes hydrostatic equilibrium between gravity and the 
internal pressure of a spherically symmetric static star: 
\begin{align}
\label{tov1}
\frac{dP}{dr} = & -\frac{\left(\varepsilon+P\right) \left(M+4\pi r^3 P\right)}{r(r-2 GM)}, \\
\frac{dm}{dr} = & 4\pi r^2 \varepsilon.
\label{tov2}
\end{align} 
The moment of inertia of a neutron star can be also calculated by using slow rotation approximation
for which the general metric describing the geometry outside the star can be written as~\cite{Hartle:1967he,Morrison:2004df} 
\begin{align}
ds^2_r = & -e^{2\nu(r)}dt^2 + e^{2\lambda(r)} dr^2 + r^2 d\theta^2 \nonumber \\
& + r^2 \sin^2\theta d\phi^2 - 2\omega(r)r^2\sin^2\theta dt d\phi ,
\label{metric}  
\end{align}
where $\omega(r)$ represents the angular velocity of the local inertial frames.
The equation for the rotational drag $\bar{\omega}(r)\equiv \Omega - \omega(r)$ of a star with angular 
frequency $\Omega$ is given by 
\begin{equation}
\frac{d}{dr}\left[r^{4}j(r)\frac{d\bar{\omega}(r)}{dr}\right]
+ 4r^{3}\frac{dj(r)}{dr}\bar{\omega}(r) = 0\;,
\label{OmegaBar}
\end{equation}
where $j(r)\equiv e^{-\nu(r)-\lambda(r)}$ is equal to $e^{-\nu(r)}\sqrt{1-2m(r)/r}$ for $r \le R$, and becomes
unity for $r > R$. For a slowly rotating neutron star with the rotation angular velocity 
$\Omega \ll \Omega_{\rm max} \approx \sqrt{M/R^3}$, the frequency $\bar{\omega}(r)$ obeys the boundary condition 
$\bar{\omega}(R)/\Omega = 1 -2I/R^3$. The total moment of inertia $I$ of the neutron star can be calculated 
from the integral~\cite{Hartle:1967he,Morrison:2004df} 
\begin{equation}
I = \frac{8\pi}{3}
\int_{0}^{R} r^{4} e^{-\nu(r)}\frac{\bar{\omega}(r)}{\Omega}
\frac{\left[\varepsilon(r)+P(r)\right]}{\sqrt{1-2m(r)/r}} dr .
\label{MomInertia}
\end{equation}
 
The dimensionless tidal deformability $\Lambda$ of a neutron star in the gravitational wave signal can be expressed in terms 
of the star's mass, radius, and the tidal Love number $k_2$. Considering quadrupolar ($l=2$), static, even-parity metric 
perturbations $h_{\alpha \beta}$ in the Regge-Wheeler gauge~\cite{Regge:1957td,Thorne67}, 
\begin{align} 
h_{\alpha\beta} = & Y_{2m}(\theta,\phi)  \nonumber \\
&  \times \text{diag}\left[e^{-\nu(r)}H_0,e^{\lambda(r)}H_2,r^2K(r),r^2\sin^2\theta K(r)\right] ,
\label{hsub}
\end{align}
the tidal Love number $k_2$ can be obtained in terms of the metric function value
and its derivative on the star’s surface, which further gives the tidal deformability parameter 
$\lambda$ as \cite{Flanagan:2007ix,Hinderer:2007mb,Hinderer:2009ca,Damour:2012yf},
\begin{eqnarray} 
\lambda = \frac{2}{3} k_2 R^5 .
\label{lam}
\end{eqnarray}
The mass normalized dimensionless value of the tidal deformability $\Lambda$ is then defined as a function 
of Love number, gravitational mass, and radius as
\begin{equation} 
\Lambda \equiv \frac{\lambda}{M^5} = \frac{2}{3} k_2 \left(\frac{R}{M} \right)^5  \equiv \frac{2}{3} C^{-5} ,
\label{Lam}
\end{equation}
where $C$ is the compactness of the star with mass $M$ and radius $R$.

\subsection{Core-crust transition properties of neutron star}

The objective of this section is to examine the characteristics of the NS core-to-crust transition properties. 
Under small-amplitude density fluctuations, one can search for the breakdown of the stability criteria of the homogeneous core 
resulting in the appearance of nuclear clusters and, subsequently, a transition to the inner crust. There are several 
ways to determine the transition density from the core side: using the Vlasov equation method~\cite{Pais:2010dp}, random phase approximation~\cite{Horowitz:2000xj,Sulaksono:2006fb}, 
dynamical method~\cite{Xu:2009vi,Ducoin:2011fy}, or thermodynamical method~\cite{Kubis:2004xh, Kubis:2006kb, Lattimer:2006xb}.

According to the thermodynamical approach, a system has to satisfy the requirements of both the mechanical and 
chemical stabilities in order to be stable against small density fluctuations i.e.~\cite{Kubis:2004xh, Kubis:2006kb, Lattimer:2006xb},
\begin{eqnarray}
\label{thermo1}
-\left (\frac{\partial P}{\partial v}\right )_{\hat{\mu}} > 0,  \\ 
-\left (\frac{\partial {\hat{\mu}}} {\partial q_c}\right )_{v} > 0.
\label{thermo2}
\end{eqnarray}
Here $P$ represents the total pressure of the neutron star matter, the volume and charge per baryon number 
are denoted by the variables $v$ and $q_c$. Further, the $\beta-$equilibrium condition of the system implies 
$\hat{\mu} \equiv \mu_n - \mu_p = \mu_e$ for the chemical potential of neutrons, protons, and electrons.
Since in Eq. (\ref{thermo1}) the derivative is carried out at a constant $\hat{\mu}$, the electron pressure 
does not affect this term. This allows one to write Eq. (\ref{thermo1}) as
\begin{equation}
-\left (\frac{\partial P_b}{\partial v}\right )_{\hat{\mu}} > 0 ,
\label{thermo3}
\end{equation}
in terms of the pressure $P_b$ due to the baryons.

For a given density $\rho$ and the isospin asymmetry $\delta$ of the $\beta$-stable system, the stability criteria 
Eqs. (\ref{thermo2}) and (\ref{thermo3}) can be represented in terms of the energy per nucleon $e(\rho, \delta)$.
Using the relations $P_b (\rho,\delta) = \rho^2 \partial e(\rho, \delta)/\partial \rho$ for baryons and 
$\hat{\mu} = \mu_n - \mu_p = 2 \partial e(\rho,\delta)/\partial \delta$,
the mechanical stability condition Eq. (\ref{thermo3}) can be expressed as~\cite{Lattimer:2006xb,Kubis:2006kb, Pais:2016nzh,Gonzalez-Boquera:2017uep}
\begin{align}
-\left( \frac{\partial P_b}{\partial v} \right)_{\hat{\mu}} = & \rho^2 \left[ 2\rho 
\frac{\partial e(\rho,\delta)}{\partial\rho} + \rho^2 \frac{\partial^2 e(\rho,\delta)}{\partial\rho^2} \right. 
\notag \\
& \left. -\frac{ \left( \rho \frac{\partial^2 e(\rho,\delta)}{\partial\rho \partial\delta} \right)^2}
{\frac{\partial^2 e(\rho,\delta)}{\partial \delta^2}} \right]  > 0 .
\label{thermo4}
\end{align}
Using the expression for the charge $q = (1- \delta)/2 - \rho_e/\rho$, the chemical stability condition Eq.~(\ref{thermo2}) 
can be written as~\cite{Lattimer:2006xb,Kubis:2006kb} 
\begin{equation}\label{thermo5}
-\left( \frac{\partial q}{\partial \hat{\mu}} \right)_v = \frac{1}{4} \left[ 
\frac{\partial^2 e(\rho,\delta)}{\partial \delta^2} \right]^{-1} + \frac{\mu_e^2}{\pi^2 \rho} > 0. 
\end{equation}
Since Eq. (\ref{thermo5}) is typically valid in neutron stars, the crust-core transition density $\rho_t$ can be obtained
by using the condition $( \partial P_b/\partial v )_{\hat{\mu}}=0$ in Eq. (\ref{thermo4}).
Further, Eqs. (\ref{EOSANM}) and (\ref{thermo4}) provide the explicit stability condition in terms of 
the density derivative of $e_0$ and $e_{\mathrm{sym}}(\rho)$ as~\cite{Lattimer:2006xb,Kubis:2006kb}
\begin{widetext}
\begin{align}
-\left( \frac{\partial P_b}{\partial v} \right)_{\hat{\mu}}  =& 
\rho^2\left[\rho ^{2}\frac{\partial^2 e_0(\rho)} {\partial\rho^2} 
+ 2\rho \frac{\partial e_0(\rho)}{\partial\rho} + \delta^2 \left( \rho^2
\frac{\partial^2 e_{\mathrm{sym}}(\rho)}{\partial \rho^2} 
+ 2\rho \frac{ \partial e_{\mathrm{sym}}(\rho)}{\partial \rho} -2e_{\mathrm{sym}}^{-1}(\rho)\left( \rho \frac{\partial e_{\mathrm{sym}}(\rho)}{\partial \rho}\right)^{2} \right) 
 \right ] > 0 .
\label{thermo6}
\end{align}
\end{widetext}
Hence, vanishing of Eq. (\ref{thermo6}) provides the crust-core transition density $\rho_t^{\rm PA}$ in the parabolic approximation of the EoS.
The first and second derivatives of the symmetric nuclear EoS $e_0(\rho)$ and symmetry energy $e_{\rm sym}(\rho)$ 
explicitly appear in the stability condition Eq. (\ref{thermo6}) that is required for stability against 
spinodal decomposition. However, due to the complexity of this relationship, it is valuable to investigate the 
dependence of the transition properties on the symmetry energy slope 
$L(\rho) \approx \partial e_{\mathrm{sym}}(\rho)/\partial \rho$, and the incompressibility of symmetric nuclear 
matter $K(\rho)$, which depends explicitly on both $\partial e_0(\rho)/\partial\rho$ 
and $\partial^{2} e_{0}(\rho)/\partial \rho^{2}$.
Note that if Eq. (\ref{thermo6}) is used to calculate the density at which the system becomes unstable, one estimates the
core-crust transition density $\rho_t^{PA}$ corresponds to the parabolic approximation (PA) of the EoS. 
Whereas, the exact value of the core-crust transition density can be obtained by using Eq. (\ref{thermo4}) for the full EoS. 
The corresponding pressure $P_t$ at the transition point can then be also extracted. 

Several model studies of pulsar glitches have established unique connection between the size of the glitch and the 
crustal moment of inertia of the NS. The glitches are sudden jumps or discontinuities in rotational
frequency in otherwise regular pulsations of rotating stars due to transfer of angular momentum within
a short time from rapidly spinning superfluid core to the crust of the pulsar \cite{Pines1985,Anderson:1975zze,Chamel:2016led}. 
The ratio between the crustal moment of inertia and the total moment of inertia $\Delta I_\mathrm{crust}/I$ 
of a neutron star can be further explored in our 
correlation analysis by using the proposed relation \cite{Lattimer:2000nx,Lattimer:2000kb,Worley:2008cb} 
\begin{align}
\frac{\Delta I_\mathrm{crust}}{I} \approx & \frac{28\pi P_t R^3}{3M} \frac{\left(1-1.67 C-0.6 C^2 \right)}{C} \nonumber \\ 
& \times \left[ 1 + \frac{2 P_t \left( 1 + 5C - 14C^2\right)}{\rho_t m C^2}\right]^{-1}, 
\label{eq:Iaprox}
\end{align}
where $C$ is the compactness of neutron star and $m$ is the mass of baryons.
This relation incorporates the mass $M$, radius $R$, as well as the transition pressure $P_t$ and/or 
transition density $\rho_t$, which strongly depend on the model neutron star EoS.

\section{Results and discussion}\label{sec:Results}

\subsection{Correlation of nuclear matter parameters with the bulk properties of NS}

%%%%%%%%%%%%%%%%%%%%%%%%%%%%%%%%%%%%%%%%%%%%
\begin{figure}[t]
\centering
 \includegraphics[width=\linewidth,angle=0]{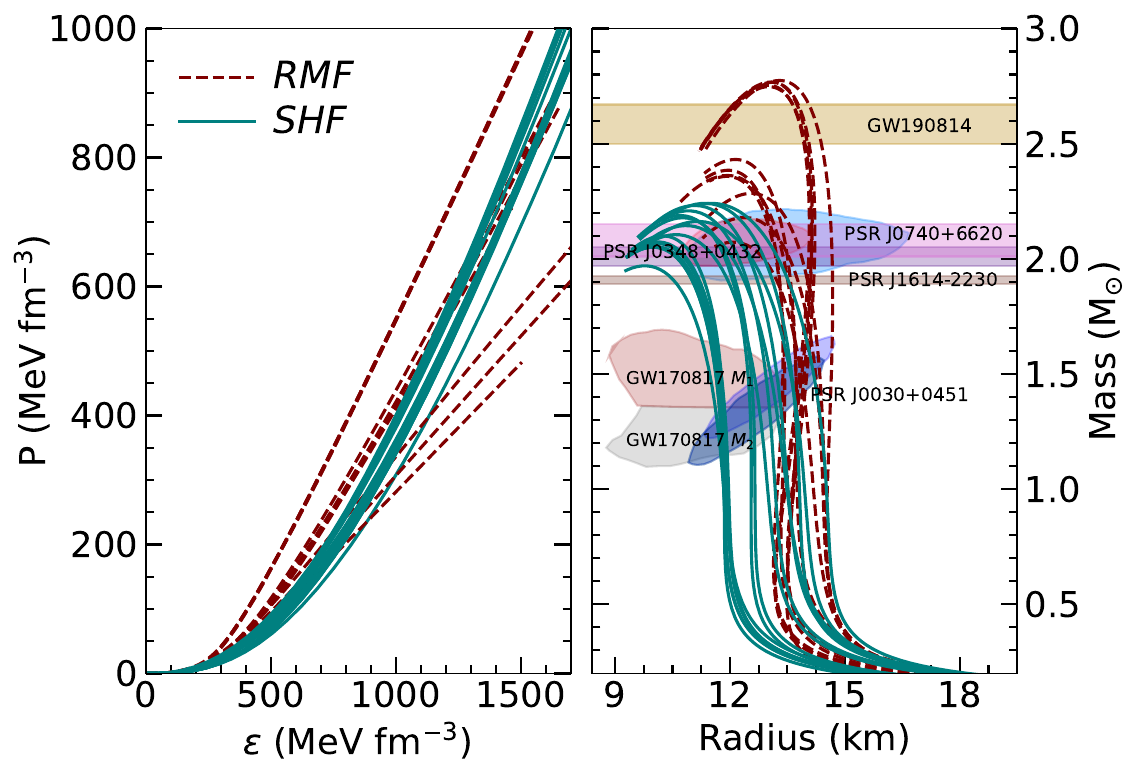}
\caption{ Pressure $P$ as a function of energy density $\varepsilon$ (left), 
and mass-radius sequences (right) for neutron stars for EoS from Skyrme–Hartree–Fock
(SHF: solid green lines) and relativistic mean-field (RMF: dashed maroon lines) models.
The contours and bands refer to
$M-R$ constraints from the NICER measurements of PSR J0030+0451~\cite{Riley2019}
and PSR J0740+6620~\cite{Riley2021}, the pulsar PSR J0348+0432~\cite{Antoniadis13}
and PSR J1614-2230~\cite{Demorest10,Arzoumanian18}, the gravitational wave GW170817 event~\cite{LIGOScientific:2017vwq}, and the secondary component of GW190814 
with mass of $2.59^{+0.08}_{-0.09}M_\odot$~\cite{LIGOScientific:2020zkf}.}
\label{fig:Pe-MR}
\end{figure}
%%%%%%%%%%%%%%%%%%%%%%%%%%%%%%%%%%%%%%%%%%%%

To facilitate visualization of the EoSs used in this work, we have displayed the pressure $P$ for 
neutron star matter (in charge neutral and $\beta$-equilibrium conditions) as a function of energy density $\varepsilon$ 
in Fig. \ref{fig:Pe-MR} (left panel). The corresponding mass–radius sequences obtained by solving the TOV equations 
(\ref{tov1}) and (\ref{tov2}) for these EoSs are also shown in Fig. \ref{fig:Pe-MR} (right panel). In general, the softer EoSs
in the Skyrme-Hartree-Fock  models (solid green lines) lead to lower maximum mass stars compared to the stiff relativistic mean-field 
models (dashed brown lines).
The EoSs employed in this study are consistent with the well-established observational constraints on the maximum mass
of neutron stars $2M_{\odot}$~\cite{Demorest10,Arzoumanian18}. Moreover, these EoSs predict radii in the range of 
$R_{1.4} \approx 11.6$–$14.6$ km for a canonical $1.4M_\odot$ neutron star, which lies well within the observed values. 
Hence, the observational bounds from radius and tidal deformability can be suitably employed to explore their correlations 
with the nuclear EOS parameters to reliably extract these parameters.

%%%%%%%%%%%%%%%%%%%%%%%%%%%%%%%%%%%%%%%%%%%%
\begin{figure}[t]
\centering
 \includegraphics[width=\linewidth,angle=0]{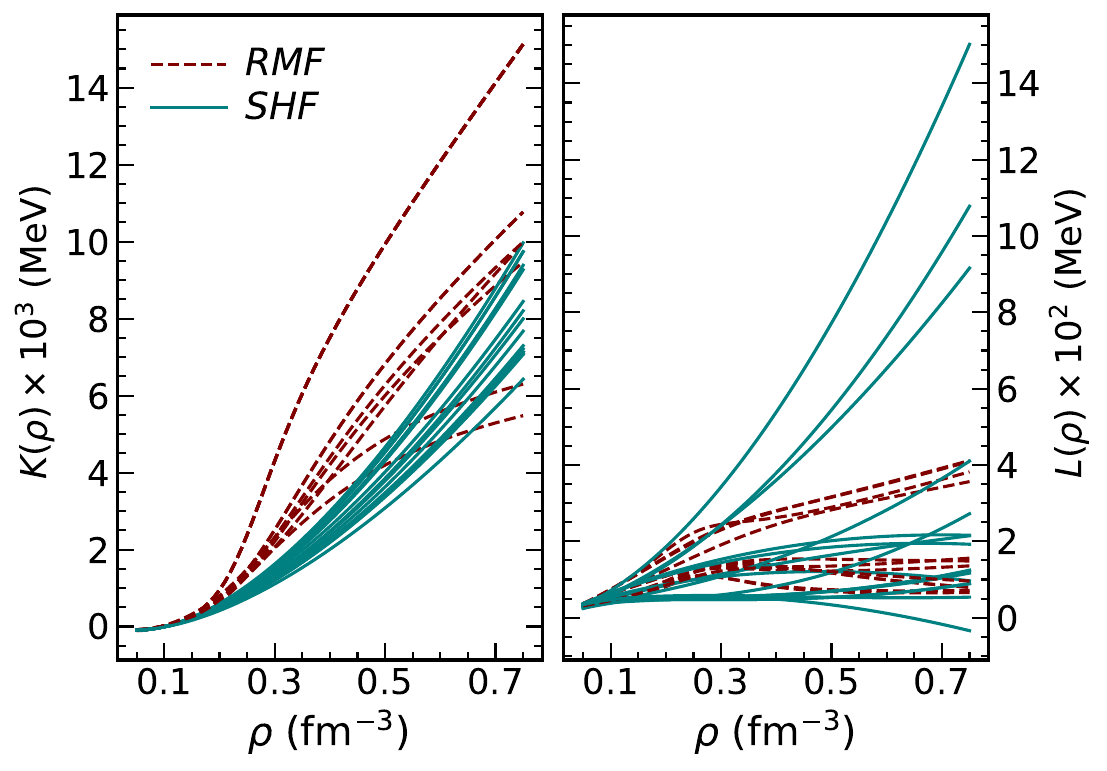}
\caption{Baryon density dependence of symmetric nuclear matter incompressibility $K(\rho)$ and 
symmetry energy slope $L(\rho)$ in the Skyrme-Hartree Fock (SHF) models (solid green lines) and the relativistic mean field (RMF) models (dashed maroon lines).}
\label{fig:LK-rho}
\end{figure}
%%%%%%%%%%%%%%%%%%%%%%%%%%%%%%%%%%%%%%%%%%%%

%%%%%%%%%%%%%%%%%%%%%%%%%%%%%%%%%%%%%%%%%%%%
\begin{figure*}[t]
\centering
\includegraphics[width=1.0\linewidth,angle=0]{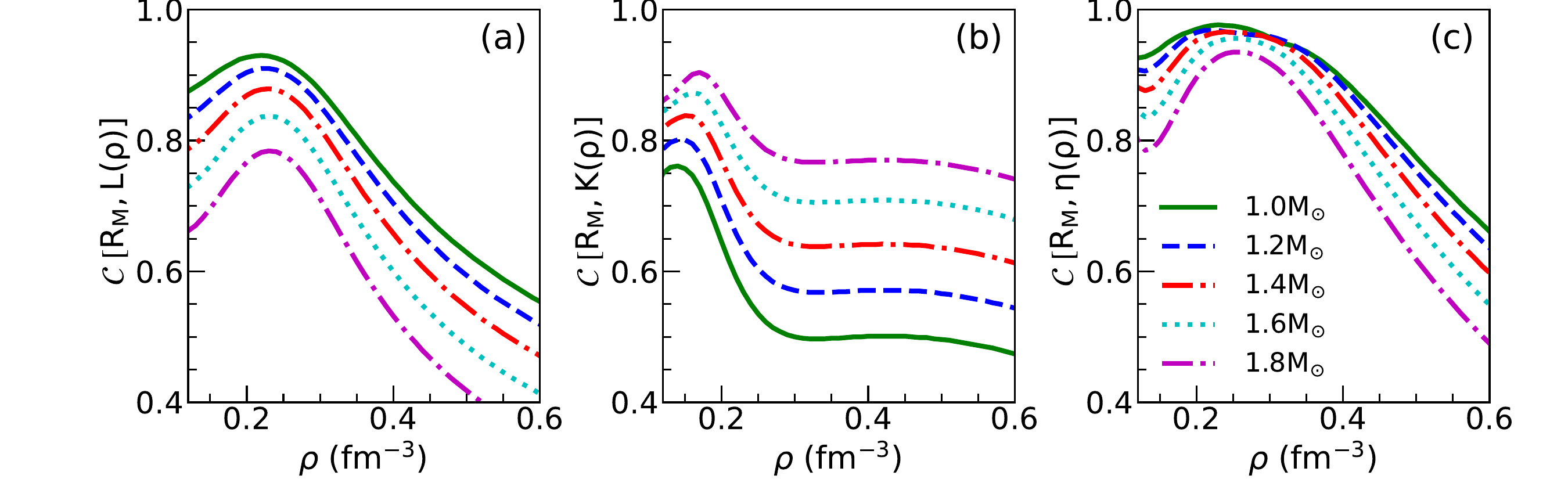}
\caption{Pearson correlation coefficient as a function of baryon density between neutron star 
radii at a fixed mass $R_M$  and $L(\rho)$, $K\rho)$, $\eta(\rho) = [K(\rho)L^2(\rho)]^{1/3}$ 
calculated using SHF and RMF model.}
\label{fig:Corr-R-LKE}
\end{figure*}
%%%%%%%%%%%%%%%%%%%%%%%%%%%%%%%%%%%%%%%%%%%%

In order to investigate the correlation between of neutron star properties with the key nuclear matter parameters, we present in 
Fig. \ref{fig:LK-rho} the variation in the incompressibility of symmetric nuclear matter $K(\rho)$ and the symmetry energy slope 
$L(\rho)$ with the baryon number density for our representative set of RMF and SHF models. These two classes of models 
exhibit opposite behavior for the EoS parameters. At supra-saturation densities, the spread in isoscalar EoS parameter 
$K(\rho)$ is larger in the RMF with $230 < K(\rho_0)/{\rm MeV} < 300$ as compared to the deviation in the relatively softer
EoS in the nonrelativistic SHF model. 
In contrast, the isovector EoS parameter $L(\rho)$ displays a much smaller variance of a few hundred of MeV only in 
RMF relative to the large spread ($\sim 1000$ MeV) seen in SHF sets at typical central densities of (3-5)$\rho_0$ in neutron stars. 

To analyze the correlation between the neutron star bulk and crustal observables with the nuclear matter (NM) parameters of the EoS,
we take recourse to Pearson correlation coefficient ${\cal C}[a,b]$ that describes quantitatively the linear correlation 
between two quantities $a$ and $b$ and can be expressed as \cite{Brandt:1976zc}
\begin{equation}
{\cal C}[a,b] = \frac{\sigma_{ab}}{\sqrt{\sigma_{aa}\sigma_{bb}}} ,
\label{eq:cc}
\end{equation}
where the covariance $\sigma_{ab}$ is given by
\begin{equation}
\sigma_{ab}=\frac{1}{N_m}\sum_i a_i b_i - \left(\frac{1}{N_m}\sum_i a_i\right)\left(\frac{1}{N_m}\sum_i b_i\right) .
\end{equation}
The summation index $i$ runs over the number of models $N_m$ used in the analysis; 
$a_i$ refers to the star properties (radius, moment of inertia, deformability, core-crust transition density)
at a fixed NS mass and $b_i$ corresponds to the nuclear matter EoS parameters ($K(\rho)$, $L(\rho)$, $\eta(\rho)$).
A correlation coefficient ${\cal C}[a,b] = \pm 1$ indicates perfect correlation/anticorrelation between the two
quantities, and ${\cal C}[a,b] = 0$ corresponds to no correlation.

Figure \ref{fig:Corr-R-LKE}(a) shows the density dependence of Pearson correlation coefficients between 
the EoS parameters $L(\rho)$ and the neutron star radius $R$ at fixed values of the star mass $M$. 
As evident from the figure, the strongest correlation of ${\cal C}[R_{1.0},L(\rho)] = 0.93$ occurs at a 
density $\rho = 0.25$ fm$^{-3}$ for the smallest star mass $M = 1.0M_\odot$ studied. While the strength of the
correlation decreases for larger mass stars, the correlation peak appears at a fixed density $\rho \approx 1.6\rho_0$
independent of the selection of the NS mass. This suggests that precision measurements of radius
of low mass stars would uniquely determine the slope of symmetry energy at densities beyond the
saturation density. In fact, correlation analysis of maximum mass stars $M_{\rm max}$ with $L(\rho_0)$
at the saturation density in the RMF model showed the strongest correlation appears for low
$M_{\rm max}$ stars. 
We also present in Fig. \ref{fig:Corr-R-LKE}(b) the correlation coefficients between $R$ and the 
symmetric NM compressibility $K(\rho)$ as a function of density. The maximum correlation 
${\cal C}[R_{1.8},K(\rho)] = 0.91$ is seen at about the same density $\rho \approx 1.6\rho_0$ as for $L(\rho)$ 
but for the massive $1.8M_\odot$ star. While the correlation function decreases and the peak shifts 
to lower densities for smaller mass stars, the correlation nearly flattens at densities $\rho > 0.30$ fm$^{-3}$ 
for stars at a fixed mass $M$.
To understand the complicated correlation of $R$ with $L(\rho)$ and $K(\rho)$, we note the NS radii is 
determined by the degenerate pressure of neutron-rich matter that supports the star against gravitational collapse. 
Indeed, an empirical relation $R \propto P^{1/4}$ was deduced at fiducial densities of $(1-2)\rho_0$ \cite{Lattimer:2006xb,Lattimer:2000nx}.
From Eqs. (\ref{EOSANM}) - (\ref{Esym}), the pressure corresponding to the symmetric NM compressibility 
and the symmetry energy slope terms is given by \cite{Alam:2016cli}
\begin{equation}
P = \frac{\rho^2}{3\rho_0} \left[ \frac{K}{3} \left(\frac{\rho}{\rho_0} - 1 \right) + L \delta^2 \right] .
\label{eq:pres}
\end{equation}
At the saturation density, only $L(\rho_0)$ contributes to pressure, and at higher densities the 
$\delta^2$ term monotonically decreases enforcing a falling contribution to $P$ from $L$.
This suggests that the radius of a NS averages out $L(\rho)$ at about $(0.5-2)\rho_0$
and causes the low mass neutron stars to be strongly correlated with $L$ as seen in Fig. \ref{fig:Corr-R-LKE}(a).
On the other hand, at larger densities $\rho > \rho_0$ the compressibility/stiffness term $K(\rho)$ increasingly dominates 
the total pressure and also generates stars that are more massive in the sequence of NS. 
The higher mass stars can then probe the pressure at higher interior densities. As a result, the 
strongest correlation ${\cal C}[R,K(\rho)]$ is seen for the massive stars at around 1.6 times the 
saturation density relative to the low mass stars.

In Fig.  \ref{fig:Corr-R-LKE}(c) we depict the density dependence of the correlation between star radius 
and the combined EoS parameter $\eta = [K L^2]^{1/3}$ which was found \cite{Sotani:2013dga,Sotani:2015lya}
to yield enhanced correlation for 
various NS observables at the saturation density.  We find the strongest correlation of 
${\cal C}[R,\eta(\rho)] \approx 0.97$ over a wider density range at around $\rho = 0.25$ fm$^{-3}$ 
relative to the correlation for the individual $K$ and $L$ parameters. Further, the strong $R-\eta$ correlation has a small
dependence on the mass of NS implying the possibility to apply stringent constraint on the EoS at
$\sim 2\rho_0$ from radius measurements only regardless of the mass of NS.

%%%%%%%%%%%%%%%%%%%%%%%%%%%%%%%%%%%%%%%%%%%%
\begin{figure}[t]
\centering
\includegraphics[width=\linewidth,angle=0]{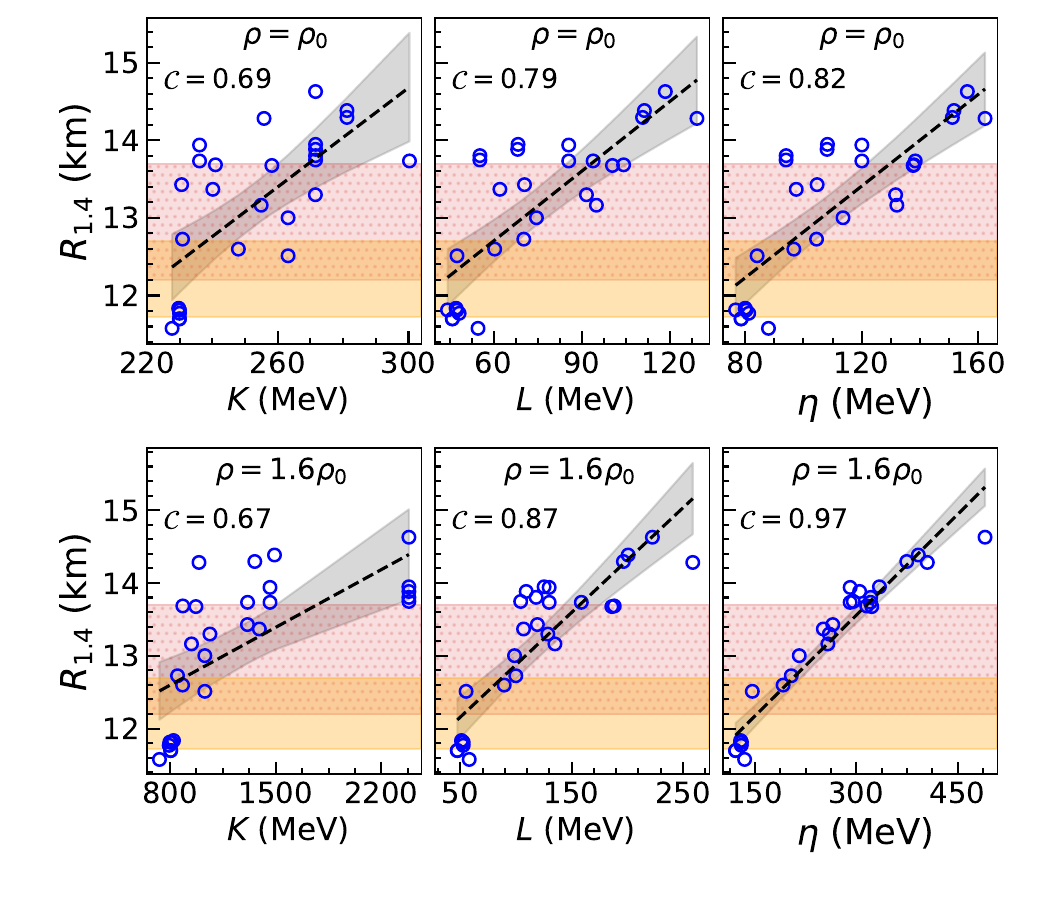}
\caption{Correlation between $R_{1.4}$ with the EoS parameters $K$, $L$ and their specific combination $\eta$ 
at baryon densities $\rho_0$ (upper panels) and $1.6\rho_0$ (lower panels) in the SHF and RMF models of EoS. The lines represent the linear best fit and the gray shaded region correspond to the 95\% confidence band.
The horizontal bands refer to radius bound of $R_{1.4} = 12.9^{+0.8}_{-0.7}$ km (magenta) from GW190814 event \cite{LIGOScientific:2020zkf} and $R_{1.4} = 12.20^{+0.50}_{-0.48}$ km (orange) estimated from analysis of 
GW170817 event and its electromagnetic counterparts plus various pulsar data \cite{Koehn:2024set}.}
\label{fig:corr-R14}
\end{figure}
%%%%%%%%%%%%%%%%%%%%%%%%%%%%%%%%%%%%%%%%%%%%

%%%%%%%%%%%%%%%%%%%%%%%%%%%%%%%%%%%%%%%%%
\setlength{\tabcolsep}{5pt}
\begin{table}[b]
\caption{\label{tab1} Coefficients of the linear fits $\mathcal{F}_{\mathcal{O}}^{\rho}$ and $\mathcal{G}_{\mathcal{O}}^{\rho}$ between the neutron star observables 
$\mathcal{O}\in \{R_{1.4},\Lambda_{1.4}\}$ and the EoS parameter $\eta =[KL^2]^{1/3}$ 
[given in Eqs. (\ref{eq:eta_R14}), (\ref{eq:eta_L14})] and the parameter $\zeta = K + \alpha L$  [given in Eqs. (\ref{eq:KL_R14}), (\ref{eq:KL_Lambda14})] at densities $\rho=\rho_0$ and $\rho = 1.6\rho_0$.}
\begin{center}
\begin{tabular}{ccccc}
\hline\hline
Correlation  & $\rho$  &  $\alpha$ & $\mathcal{F}_{\mathcal{O}}^{\rho}$ & $\mathcal{G}_{\mathcal{O}}^{\rho}$ \\
\hline
$R_{1.4}$--$\eta$         &  $\rho_0$        &     --  &   $0.03 \pm 0.00$  & $9.89 \pm 0.45$ \\
$R_{1.4}$--$\eta$         &  $1.6\rho_0$     &     --  &   $0.01 \pm 0.00$  & $10.79 \pm 0.13$ \\
$\Lambda_{1.4}$--$\eta$   &  $\rho_0$        &     --  &   $8.50 \pm 1.42$  & $-237.31 \pm 161.82$ \\
$\Lambda_{1.4}$--$\eta$    &  $1.6\rho_0$      &     --  &   $2.89 \pm 0.14$  & $-31.41 \pm 38.78$ \\
$R_{1.4}$--$\zeta$        &  $\rho_0$         &     1.23 &   $1.82 \pm 0.20$  & $6.89 \pm 0.73$\\
$R_{1.4}$--$\zeta$         &  $1.6\rho_0$     &    16.44 &   $0.07 \pm 0.00$  & $10.78 \pm 0.14$ \\
$\Lambda_{1.4}$--$\zeta$   &  $\rho_0$        &     0.84  &   $6.81 \pm 0.89$  & $-1435.37 \pm 281.95$ \\
$\Lambda_{1.4}$--$\zeta$   &  $1.6\rho_0$      &    11.45  &   $0.29 \pm 0.01$  & $-65.88 \pm 26.89$ \\
\hline\hline
\end{tabular}
\end{center}
\end{table}
%%%%%%%%%%%%%%%%%%%%%%%%%%%%%%%%%%%%%%%%%

Figure \ref{fig:corr-R14} shows the correlation between the radius $R_{1.4}$ of a canonical $1.4M_\odot$ neutron star 
and the slope $L$, compressibility $K$ 
and their combination $\eta$ at $\rho_0$ and $1.6\rho_0$. While moderate correlations are seen for
$L$ and $K$, a much stronger correlation is prevalent in the combined EoS parameter $\eta$.
More importantly $R_{1.4}$ and $\eta(1.6\rho_0)$ are distinctly well correlated at $\rho=1.6\rho_0$
given that a diverse class of EoS are employed in the analysis. 
Also shown are the linear regression constructed between the radius and the EoS parameters with
95\% confidence band by considering the scatters in the EoS. For the EoS parameter $\eta$ this gives at $\rho=\rho_0$ and $\rho=1.6\rho_0$
\begin{align}
\frac{R_{1.4}}{\rm km} = & \mathcal{F}_{R_{1.4}}^{\rho}\frac{\eta(\rho)}{\rm MeV} + \mathcal{G}_{R_{1.4}}^{\rho},
\label{eq:eta_R14}
\end{align}
where $\mathcal{F}_{R_{1.4}}^{\rho}$ and $\mathcal{G}_{R_{1.4}}^{\rho}$ are the fit parameters at the densities as listed in Table \ref{tab1}.
Constraints on the EoS parameters can be imposed by relating them to the measured radius of a canonical neutron star, from the gravitational wave event GW170817 \cite{LIGOScientific:2017vwq,De:2018uhw,Koehn:2024set}, as well as from the 
more recent detection of the secondary component in GW190814 \cite{LIGOScientific:2020zkf}.

%%%%%%%%%%%%%%%%%%%%%%%%%%%%%%%%%%%%%%%%%
\setlength{\tabcolsep}{5pt} 
\begin{table*}[t]
\caption{\label{tab2} Neutron star radius $R_{1.4}$ (in km), tidal deformability $\Lambda_{1.4}$, and the corresponding estimated nuclear matter parameters $\eta=[K L^2]^{1/3}$, incompressibility $K$, and the symmetry energy slope $L$ 
(in MeV unit) at densities $\rho_0$ and $1.6\rho_0$.}
\vspace*{-.3cm}
\begin{center}
\begin{tabular}{lccccccc}
\hline\hline
GW event & NS bounds & $\eta(\rho_0)$ & $K(\rho_0)$ & $L(\rho_0)$ & $\eta(1.6\rho_0)$ & $K(1.6\rho_0)$ & $L(1.6\rho_0)$ \\
\hline \\ [-2mm]
GW170817 \cite{LIGOScientific:2017pwl}            &  $\Lambda_{1.4}=190^{+390}_{-120}$     &    $50.3^{+45.9}_{-14.1}$  &   $221.4^{+27.3}_{-10.1}$  & $24.0^{+35.8}_{-9.0}$ &$76.6^{+134.9}_{-41.5}$ & $97.5^{+267.8}_{-67.5}$ & $67.9^{+93.1}_{-29.9}$ \\[1mm]
GW170817+EM+PSR~\cite{Koehn:2024set}              &  $R_{1.4}=12.20^{+0.50}_{-0.48}$        &   $79.2^{+16.9}_{-16.3}$ &   $235.2^{+9.9}_{-10.8}$  & $46.0^{+14.3}_{-12.6}$ &$153.5^{+54.4}_{-52.2}$&$342.4^{+124.8}_{-119.8}$& $102.8^{+35.9}_{-34.4}$\\[1mm]
\hline
GW190814 \cite{LIGOScientific:2020zkf} &   $\Lambda_{1.4}=616^{+273}_{-158}$    &    $100.4^{+32.1}_{-18.6}$ &   $250.8^{+15.0}_{-10.0}$  & $63.5^{+30.1}_{-15.8}$ &  $224.0^{+94.5}_{-54.7}$& $391.5^{+201.6}_{-113.6}$ & $169.5^{+63.9}_{-37.3}$\\[1mm]
GW190814 \cite{LIGOScientific:2020zkf}         &   $R_{1.4}=12.9^{+0.8}_{-0.7}$       &     $102.9^{+27.1}_{-23.7}$  &   $248.7^{+12.5}_{-13.5}$  & $66.2^{+25.6}_{-20.3}$ &$229.6^{+87.0}_{-76.1}$ & $517.2^{+199.8}_{-174.8}$ & $153.0^{+57.4}_{-50.2}$\\ [1mm]   
GW190814: $\Lambda_{1.4}$--$R_{1.4}$         &   $R_{1.4}=13.03^{+0.77}_{-0.59}$       &     $107.3^{+26.1}_{-20.0}$  &   $250.9^{+11.6}_{-10.8}$  & $70.2^{+24.9}_{-17.5}$ &$243.8^{+83.7}_{-64.1}$ 
 & $549.6^{+192.4}_{-147.4}$&$162.3^{+55.2}_{-42.3}$\\[1mm]              
\hline\hline
\end{tabular}
\end{center}
\end{table*}
%%%%%%%%%%%%%%%%%%%%%%%%%%%%%%%%%%%%%%%%%

%%%%%%%%%%%%%%%%%%%%%%%%%%%%%%%%%%%%%%%%%%%%
\begin{figure}[t]
\centering
\includegraphics[width=.9\linewidth,angle=0]{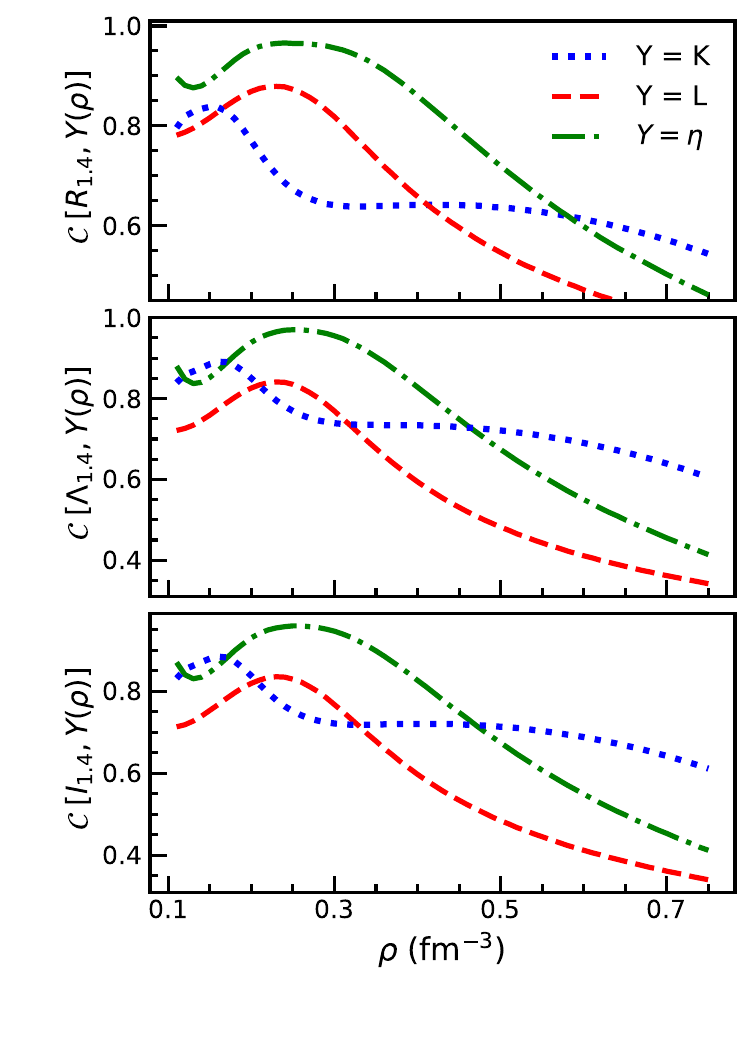}
\caption{ Density dependence of correlation coefficient between the EoS parameters
$L(\rho)$, $K\rho)$, $\eta(\rho)$ with radius $R_{1.4}$ (top panel), tidal deformability 
$\Lambda_{1.4}$ (middle panel) and moment of inertia $I_{1.4}$ (bottom panel) 
of neutron star of mass $M=1.4M_\odot$.}
\label{fig:Corr-RLam}
\end{figure}
%%%%%%%%%%%%%%%%%%%%%%%%%%%%%%%%%%%%%%%%%%%%

We compare in Fig. \ref{fig:Corr-RLam} the correlation coefficient between the dimensionless tidal deformability 
$\Lambda \propto (R/M)^5$, moment of inertia $I_{1.4}$ for $1.4M_\odot$ stars and the EoS parameters 
$L(\rho), ~ K(\rho), ~ \eta(\rho)$. 
The correlation coefficients show nearly identical density dependence behavior 
as in $R_{1.4}$. This essentially
stems from the fact that the NS compactness parameter $C$ is strictly related to bulk properties of NS such as the radius etc.

%%%%%%%%%%%%%%%%%%%%%%%%%%%%%%%%%%%%%%%%%%%%
\begin{figure}[b]
\centering
\includegraphics[width=\linewidth,angle=0]{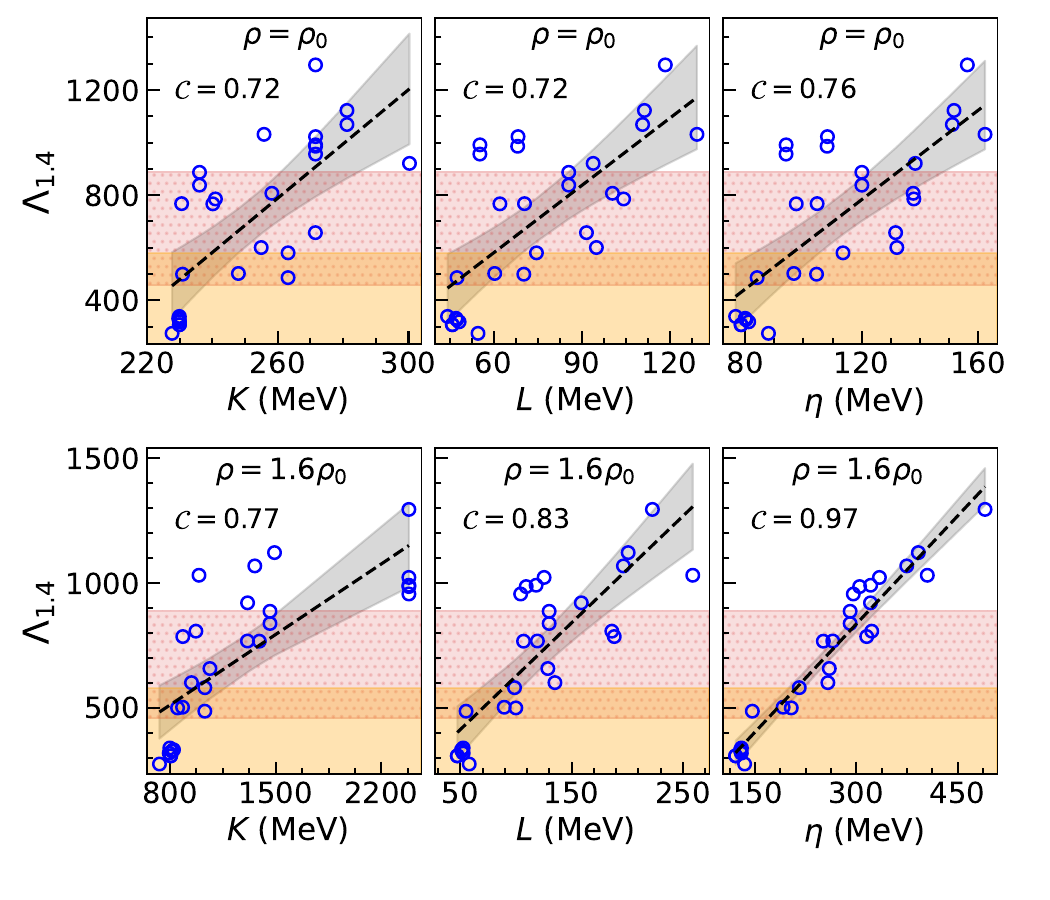}
\caption{Correlation between $\Lambda_{1.4}$ with $K$, $L$, $\eta$ 
at $\rho_0$ (upper panels) and $1.6\rho_0$ (lower panels) in the SHF and RMF models. The lines represent
the linear best fit and the gray shaded region correspond to the 95\% confidence band.
The horizontal bands refer to GW170817 \cite{LIGOScientific:2018cki} and GW190814 
\cite{LIGOScientific:2020zkf}
tidal deformability bounds of $\Lambda_{1.4} = 190^{+390}_{-120}$ (orange) and $\Lambda_{1.4} = 616^{+273}_{-158}$ (magenta), respectively.} 
\label{fig:corr-L14}
\end{figure}
%%%%%%%%%%%%%%%%%%%%%%%%%%%%%%%%%%%%%%%%%%%%

The corresponding correlation between  $\Lambda_{1.4}$ with the EoS parameters in Fig.
\ref{fig:corr-L14} also suggests stronger correlation and less cluttering at a density $\rho = 0.25$ fm$^{-3}$.
The simple mathematical relation between  $\Lambda_{1.4}$ and $\eta(\rho)$ 
can be expressed through linear regression at $\rho=\rho_0$ and $\rho=1.6\rho_0$ as 
\begin{align}
\Lambda_{1.4} = & \mathcal{F}_{\Lambda_{1.4}}^{\rho}\frac{\eta(\rho)}{\rm MeV} + \mathcal{G}_{\Lambda_{1.4}}^{\rho},
\label{eq:eta_L14}
\end{align}
where $\mathcal{F}_{\Lambda_{1.4}}^{\rho}$
and $\mathcal{G}_{\Lambda_{1.4}}^{\rho}$
are the parameters fitted to $\Lambda_{1.4}$ at the two densities and listed in Table \ref{tab1}.
The values of $\eta$ obtained from these equations can serve as an additional crucial ingredient in the fitting 
protocols of an EoS, together with other nuclear matter parameters and the finite nuclei properties, 
to optimize the model parameters in order 
to ensure that the resulting EoS is consistent with all these observational constraints. 

We will now impose the gravitational wave bounds on the explored tight correlations to estimate the nuclear EoS parameters. The GW190814 event \cite{LIGOScientific:2020zkf} involves the merger of a
massive black hole (BH) of mass $(22{-}24)\,M_\odot$ and a secondary 
component of mass about $(2.5{-}2.6)\,M_\odot$, which can be either a neutron star (NS) or
a black hole. 
Considering a NS-BH scenario for GW190814, an unique observational bound simultaneously for the tidal deformability and radius was given \cite{LIGOScientific:2020zkf} for a canonical $1.4M_\odot$ neutron star
of $\Lambda_{1.4} = 616^{+273}_{-158}$ and  $R_{1.4} = 12.9^{+0.8}_{-0.7}$ km at 90\% credible level.
Based on this radius constraint of GW190814, our correlation analysis involving radius and EoS
parameter $\eta$ (in Fig. \ref{fig:corr-R14} and Eq. (\ref{eq:eta_R14})) suggests a bound on the value of $\eta$ (in units of MeV) of $79.2 \lesssim \eta(\rho_0) \lesssim 130.0$ at the nuclear saturation density, and a more reliable higher density EoS bound $153.5 \lesssim \eta(1.6\rho_0) \lesssim 316.6$ at $1.6$ times the saturation density. While on applying the observational bound of 
$\Lambda_{1.4} = 616^{+273}_{-158}$ from GW190814 \cite{LIGOScientific:2020zkf} 
(in Fig. \ref{fig:corr-L14} and Eq. (\ref{eq:eta_L14}))
translates to central $\eta$ bound of $81.8 \lesssim \eta(\rho_0) \lesssim 132.5$ 
at the saturation density and a more restrictive bound  
of $169.3 \lesssim \eta(1.6\rho_0) \lesssim 318.5$ at $\rho = 1.6\rho_0$.
For orientation and subsequent comparison with 
different observables and analysis procedures, these results have been listed in Table \ref{tab2}.

\begin{figure}[t!]
\centering
\includegraphics[width=\linewidth,angle=0]{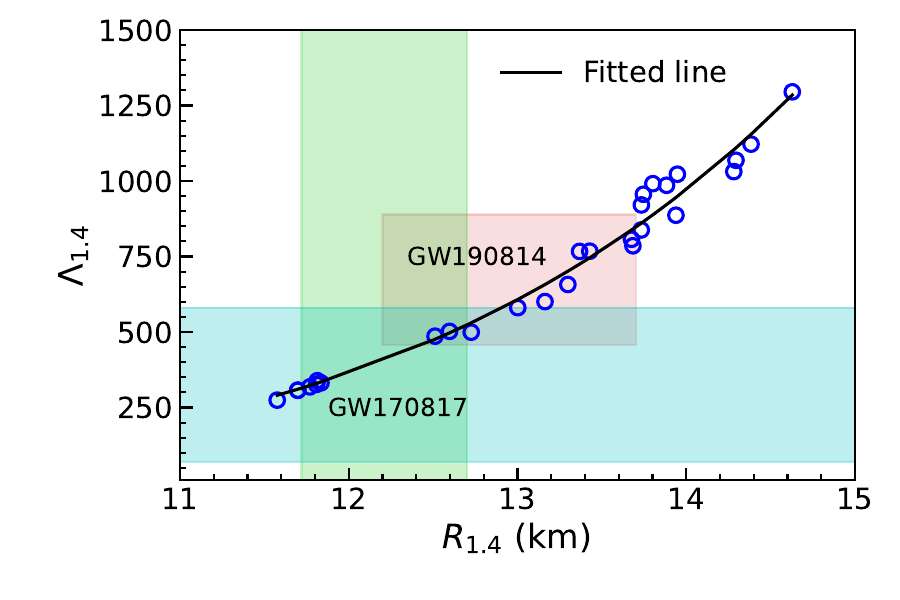}
\caption{Correlation between tidal deformability $\Lambda_{1.4}$ and radius $R_{1.4}$ of mass $1.4M_\odot$
neutron stars in the RMF and SHF models. The fitted line is represented by $\Lambda_{1.4} = 5.10 \times 10^{-5} \left(R_{1.4}/{\rm km}\right)^{6.35}$. The blue horizontal band refer to $\Lambda_{1.4} = 190^{+390}_{-120}$ from GW170817 event \cite{LIGOScientific:2018cki} and  the green vertical band refer to estimated radius 
$R_{1.4} = 12.20^{+0.50}_{-0.48}$ km from GW170817+EM+PSR \cite{Koehn:2024set}. The magenta shaded region refer to $\Lambda_{1.4} = 616^{+273}_{-158}$ and $R_{1.4} = 12.9^{+0.8}_{-0.7}$ km bounds 
from GW190814 event \cite{LIGOScientific:2020zkf}.}
\label{fig:corr-R-L}
\end{figure}

We note from Table \ref{tab2} that the bounds estimated separately from radius and tidal 
deformability constraints for GW190814 are consistent to each other for both the densities suggesting the 
robustness of our analysis procedure and reflecting the rather strong power-law correlation between the
radius and tidal deformability in the models \cite{Nandi:2018ami,Alam:2023grx}.  
Indeed from Fig. \ref{fig:corr-R-L} we find a tight correlation between $\Lambda_{1.4}$ and $R_{1.4}$
signifying an approximate universal relation of the computed EoSs which can be expressed as
$\Lambda_{1.4} = 5.10 \times 10^{-5} \left(R_{1.4}/{\rm km}\right)^{6.35}$ 
\cite{Annala:2017llu,Fattoyev:2017jql,Malik:2018zcf,Nandi:2018ami,Tsang:2019vxn,Alam:2023grx,Koliogiannis:2024okf}.
Alternatively, the extracted radius can be used to constrain the nuclear matter parameters.
In fact, the $\Lambda_{1.4} = 616^{+273}_{-158}$ bound from GW190814 in conjunction with the 
deduced power-law relation enforces a quite similar values of radius $R_{1.4}=13.03^{+0.77}_{-0.59}$ km 
and the corresponding nuclear matter parameters as seen in Table \ref{tab2}.

On the other hand, the detected GW170817 event has been conclusively proven to originate from the merger of binary neutron stars and thus could be more reliably employed to estimate the nuclear matter parameters. The initial estimate of the tidal deformability for a canonical neutron star was constrained to be $\Lambda_{1.4} < 800$ \cite{LIGOScientific:2017vwq}. 
Several studies analyzing the GW170817 data suggested that the corresponding radius is bounded 
from above by $R_{1.4} \lesssim 13.5$ km \cite{Fattoyev:2017jql,Annala:2017llu,Most:2018hfd}.
An improved estimate of $\Lambda_{1.4}$ using a low-spin prior yields $\Lambda_{1.4} = 190^{+390}_{-120}$ \cite{LIGOScientific:2018cki}
which has been commonly used in several model analysis studies, and depicted in Fig. \ref{fig:corr-L14}.
Using the tidal deformability constraint of the GW170817 event yields a bound on the $\eta$ parameter of
$36.2 \lesssim \eta(\rho_0) \lesssim 96.2$ and $35.1 \lesssim \eta(1.6\rho_0) \lesssim 211.6$ 
(shown in Table \ref{tab2}). 
Further, the power-law relation of Fig. \ref{fig:corr-R-L} along with $\Lambda_{1.4} = 190^{+390}_{-120}$ 
translates to a radius bound of $R_{1.4} = 10.83^{+2.08}_{-1.58}$ km for GW170817.
We note that the extracted $R_{1.4}$ is similar to the estimate from Bayesian analysis
of the GW event \cite{De:2018uhw}.
Evidently, the smaller tidal deformability and radius in GW170817 as compared to GW190814 result in a lower 
estimated value of $\eta$ (and other nuclear EoS parameters) as can be seen in Table \ref{tab2}.

%%%%%%%%%%%%%%%%%%%%%%%%%%%%%%%%%%%%%%%%%%%%
%%%%%%%%%%%%%%%%%%%%%%%%%%%%%%%%%%%%%%%%%%%%

Subsequent refined analysis, combining gravitational wave observations, massive pulsars
\cite{Antoniadis13,Demorest10,Arzoumanian18} and simultaneous mass-radius constraints of pulsars from NICER 
measurements \cite{Riley2019,Riley2021}, have placed important constraints on the $R_{1.4}$ estimate. 
For instance, the
combined analysis of GW170817 and its electromagnetic (EM) counterparts corresponding to kilonova AT2017gfo \cite{LIGOScientific:2017pwl,Coulter:2017wya,Lipunov:2017dwd,Shappee:2017zly,Tanvir:2017pws}, and the short 
gamma-ray burst GRB170817 \cite{LIGOScientific:2017zic,Goldstein:2017mmi,Savchenko:2017ffs} reports 
$R_{1.4} = 11.98^{+0.35}_{-0.40}$ km \cite{Pang:2022rzc}. In a recent comprehensive analysis that includes
improved models for tidal waveforms and kilonova light curves along with pulsar (PSR) observations from NICER 
lead to a more robust value of $R_{1.4} = 12.20^{+0.50}_{-0.48}$ km \cite{Koehn:2024set}.
Using this radius estimate, the GW170817+EM+PSR bound on $\eta$ turns out to be
$62.9 \lesssim \eta(\rho_0) \lesssim 96.1$ and $101.4 \lesssim \eta(1.6\rho_0) \lesssim 207.9$
which are listed in Table \ref{tab2}.

%%%%%%%%%%%%%%%%%%%%%%%%%%%%%%%%%%%%%%%%%%%%
\begin{figure}[t]
\centering
\includegraphics[width=\linewidth,angle=0]{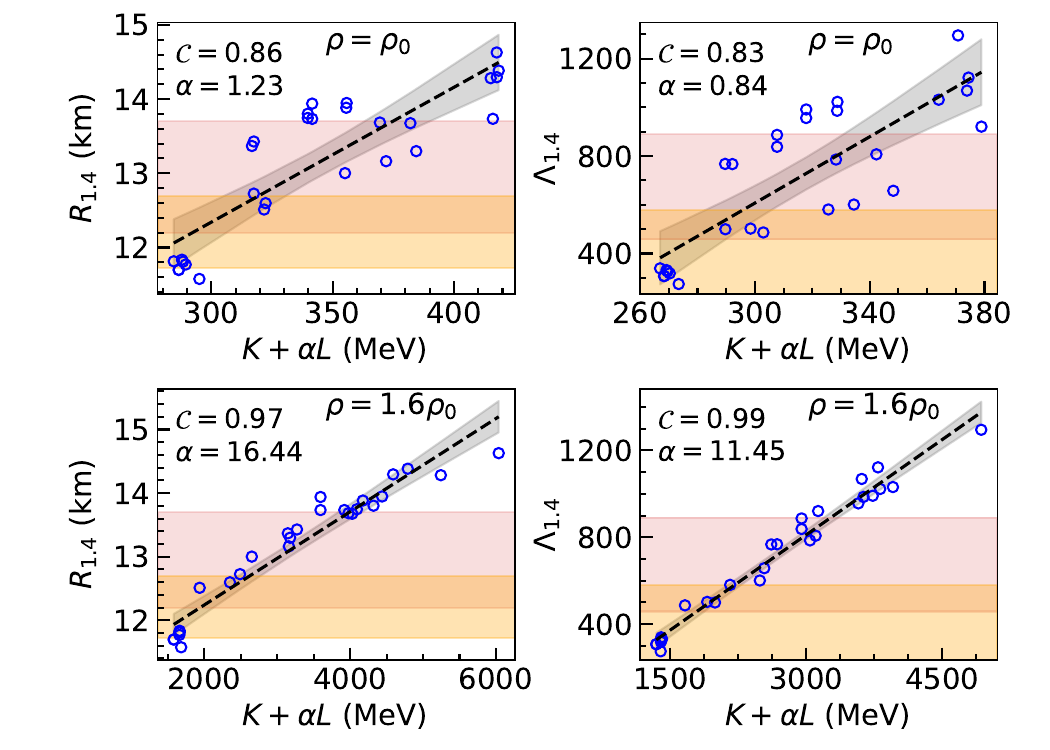}
\caption{Correlation between linear combination of nuclear matter parameters $K+\alpha L$ and $R_{1.4}$, $\Lambda_{1.4}$ at $\rho_0$ (upper panels) and $1.6\rho_0$ (lower panels). 
The horizontal shaded bounds on $R_{1.4}$ and $\Lambda_{1.4}$ are the same as in Figs. \ref{fig:corr-R14} and \ref{fig:corr-L14}.}
\label{fig:Kn-RL}
\end{figure}
%%%%%%%%%%%%%%%%%%%%%%%%%%%%%%%%%%%%%%%%%%%%

In previous studies \cite{Alam:2016cli, Malik:2018zcf} a strong correlation was demonstrated between the neutron star properties 
and linear combinations of $K$ and $L$ at the nuclear matter saturation density.
Given that the present analysis suggests a stronger correlation at a density of $1.6\rho_0$, we 
extend the analysis at this higher density by examining their correlation with the NS radius and tidal deformability. 
Figure \ref{fig:Kn-RL} depicts $R_{1.4}$ and  $\Lambda_{1.4}$ as a function of the linear combination $K + \alpha L$, 
computed at densities $\rho_0$ (top panel) and at $1.6\rho_0$ (bottom panels). In each of this analysis, the parameter 
$\alpha$ is adjusted to maximize the correlation following the approach in Refs. \cite{Alam:2016cli,Malik:2018zcf}. 
Although $L$ and $K$ are not very well correlated individually (see Figs.  \ref{fig:corr-R14} and \ref{fig:corr-L14}) 
their combination $K + \alpha L$ inject quite strong correlations particularly at the high density
at $\rho = 1.6\rho_0$ where the correlation coefficient being close to unity. 
Further, we note that the slope $L(\rho)$ has a stronger correlation at the high density
(as discussed related to Eq. (\ref{eq:pres})) which enforces a much larger weight factor than $\alpha \sim 1$
deduced at $\rho_0$. The linear regression at the saturation density $\rho_0$ depicted by solid lines in Fig. \ref{fig:Kn-RL} yields
\begin{align} \label{eq:KL_R14}
R_{1.4} = &  \mathcal{F}_{R_{1.4}}^{\rho}\frac{(K(\rho) + \alpha \: L(\rho))}{100 \:\rm MeV} + \mathcal{G}_{R_{1.4}}^{\rho} , \\
\Lambda_{1.4} =&  \mathcal{F}_{\Lambda_{1.4}}^{\rho} \frac{(K(\rho) + \alpha \: L(\rho))}{\rm MeV} + \mathcal{G}_{\Lambda_{1.4}}^{\rho},
\label{eq:KL_Lambda14}
\end{align}
where the parameters are given in Table \ref{tab1}.
We recall that due to power-law correlation between $\Lambda_{1.4}-R_{1.4}$, the analysis provides similar results at the saturation density irrespective 
of the use of $R_{1.4} = 12.9^{+0.8}_{-0.7}$ km, and $\Lambda_{1.4} = 616^{+273}_{-158}$ 
bounds from GW190814 event. For instance, combining Eqs. (\ref{eq:eta_R14}), 
(\ref{eq:KL_R14}) for the radius or Eqs. (\ref{eq:eta_L14}), (\ref{eq:KL_Lambda14}) for 
the tidal deformability, we extract nearly identical values
of incompressibility $K(\rho_0)\approx 249$ MeV and symmetry energy slope 
$L(\rho_0) \approx 65$ MeV.
Note that these estimated values are also consistent with the fiducial 
value of  $K(\rho_0)= 240\pm 20$ MeV \cite{Colo:2004mj,Todd-Rutel:2005yzo,Colo:2013yta} and $L(\rho_0)=30-87$ MeV \cite{Burgio:2021bzy,Tews:2016jhi, Zhang:2017ncy}.
Using Eqs. (\ref{eq:eta_R14}) and (\ref{eq:KL_R14}), along with the GW190814 radius constraint $R_{1.4} = 12.9^{+0.8}_{-0.7}$ km, we obtain a central value of $K(1.6\rho_0) = 517.2$ MeV and $L(1.6\rho_0) = 153.0$ MeV. 
Similarly, by applying Eqs. (\ref{eq:eta_L14}) and (\ref{eq:KL_Lambda14}) and the constraint $\Lambda_{1.4} = 616^{+273}_{-158}$, we obtain the central value $K(1.6\rho_0) = 391.5$ MeV and $L(1.6\rho_0) = 169.5$ MeV. 

On the other hand, the tidal deformability bound $\Lambda_{1.4} = 190^{+390}_{-120}$ from binary neutron star merger GW170817 event \cite{LIGOScientific:2018cki}
yields central values of $K(1.6\rho_0) = 97.5$ MeV and $L(1.6\rho_0) = 67.9$ MeV. Whereas, 
the radius bound $R_{1.4} = 12.20^{+0.50}_{-0.48}$ extracted independently
from improved models along with electromagnetic counterparts and pulsar data
GW170817+EM+PSR \cite{Koehn:2024set} provides central estimates 
of $K(1.6\rho_0) = 342.4$ MeV and $L(1.6\rho_0) = 102.8$ MeV. We note
that these estimates have consistently smaller values than those obtained by using GW190814 bounds.

\subsection{Correlation of NM parameters with NS crustal properties}

Accurate determination of the properties of neutron star crust is of paramount importance in determining
the bulk NS properties such as the radius, moment of inertia, as well as pulsar glitches, thermal evolution 
of NS in X-ray binaries \cite{Link:1999ca,Espinoza:2011pq,Andersson:2012iu}. Although the metric function necessary 
to calculate the Love number and tidal deformability from gravitational wave emission remains almost consistent 
with and without the crust \cite{Perot:2020gux}, 
discrepancies are observed primarily due to variation in the stellar radius. In Table \ref{tab3} we have listed 
the crust-core transition density $\rho_t$ and the corresponding pressure $P_t$ calculated for the models employed
in our correlation analysis. The transition density is obtained using the thermodynamic method which considers the 
full EoS i.e. Eq. (\ref{thermo6}), as well as in the approximate parabolic EoS as input using Eq. (\ref{thermo4}).
Considerable sensitivity to the EoS for $\rho_t$ is revealed: Apparently models with large slope parameter $L$ 
injects large symmetry pressure and generates crusts with small mass at a lower transition density. 
In comparison to the parabolic approximation (PA), the exact expression of the EoS yield lower values of $\rho_t$ and $P_t$.

%%%%%%%%%%%%%%%%%%%%%%%%%%%%%%%%%%%%%%%%%
\setlength{\tabcolsep}{8pt}
\begin{table}[t]
\caption{\label{tab3} The slope of symmetry energy $L$ (in MeV) at saturation density $\rho_0$ in the RMF and SHF models. 
The crust-core transition density (in fm$^{-3}$) and pressure (in units of MeV fm$^{-3}$)
calculated using exact expression for full EoS ($\rho_t^{EE}$, $P_t^{EE}$), and in the
parabolic approximation to the EoS ($\rho_t^{PA}$, $P_t^{PA}$).}
\begin{center}
\begin{tabular}{cccccc}
\hline\hline
Model & $L(\rho_0)$ & $\rho_t^{EE}$ & $P_t^{EE}$ & $\rho_t^{PA}$ & $P_t^{PA}$ \\
\hline
BSR2                  &        62.1      &     0.076  &   0.356  &  0.080 &  0.411  \\
BSR3                  &        70.5      &     0.075  &   0.422  &  0.080 &  0.505  \\
BSR6                  &        85.6      &     0.055  &   0.287  &  0.088 &  0.924  \\
FSU2                  &        85.6      &     0.069  &   0.534  &  0.086 &  1.003  \\
GM1                   &        93.9      &     0.078  &   0.483  &  0.093 &  0.861  \\
NL3                   &       118.5      &     0.067  &   0.477  &  0.086 &  0.987  \\
NL3${\sigma \rho 2}$  &        55.3      &     0.088  &   0.465  &  0.091 &  0.503  \\
NL3${\sigma \rho 3}$  &        68.3      &     0.052  &   0.217  &  0.092 &  0.790  \\
NL3${\omega \rho 2}$  &        68.2      &     0.055  &   0.254  &  0.094 &  0.891  \\
NL3${\omega \rho 3}$  &        55.3      &     0.092  &   0.630  &  0.096 &  0.687  \\
TM1                   &       110.7      &     0.069  &   0.483  &  0.087 &  0.988  \\
TM1-2                 &       111.4      &     0.069  &   0.491  &  0.087 &  0.987  \\
KDE0v1                &        54.7      &     0.089  &   0.546  &  0.096 &  0.665  \\
SK255                 &        95.0      &     0.078  &   0.407  &  0.095 &  1.015  \\
SK272                 &        91.6      &     0.081  &   0.465  &  0.094 &  0.954  \\
SKa                   &        74.6      &     0.079  &   0.407  &  0.093 &  0.791  \\
SKb                   &        47.5      &     0.078  &   0.387  &  0.094 &  0.499  \\
SkI2                  &       104.3      &     0.063  &   0.272  &  0.090 &  0.769  \\
SkI3                  &       100.5      &     0.071  &   0.327  &  0.086 &  0.584  \\
SkI4                  &        60.3      &     0.081  &   0.413  &  0.091 &  0.522  \\
SkI5                  &       129.3      &     0.060  &   0.247  &  0.089 &  0.851  \\
SkMP                  &        70.3      &     0.071  &   0.333  &  0.091 &  0.672  \\
Sly4                  &        45.9      &     0.089  &   0.486  &  0.094 &  0.548  \\
Sly5                  &        48.2      &     0.088  &   0.483  &  0.094 &  0.555  \\
Sly6                  &        47.4      &     0.087  &   0.478  &  0.093 &  0.545  \\
Sly7                  &        47.2      &     0.087  &   0.476  &  0.093 &  0.542  \\
Sly230a               &        44.3      &     0.089  &   0.455  &  0.094 &  0.487  \\
Sly230b               &        45.9      &     0.089  &   0.486  &  0.094 &  0.548  \\
\hline\hline
\end{tabular}
\end{center}
\end{table}
%%%%%%%%%%%%%%%%%%%%%%%%%%%%%%%%%%%%%%%%%

%%%%%%%%%%%%%%%%%%%%%%%%%%%%%%%%%%%%%%%%%%%%
\begin{figure}[t]
\centering
\includegraphics[width=0.9\linewidth,angle=0]{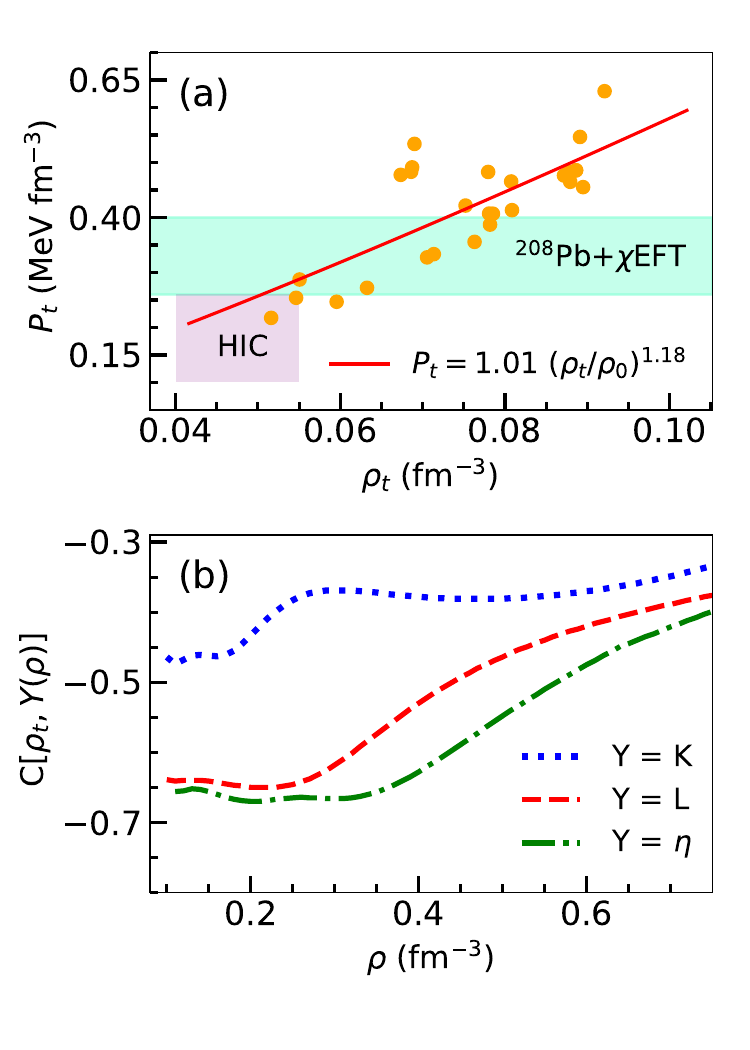}
\caption{Top panel: Core-crust transition density $\rho_t$ versus transition pressure $P_t$ 
and their correlation fitted as $P_t = 1.01 (\rho_t/\rho_0)^{1.18}$ in units of MeV fm$^{-3}$.
The bounds are from isospin diffusion bound from heavy ion collision (HIC: magenta box) \cite{Xu:2009vi}, and Bayesian 
analysis with $^{208}$Pb skin and $\chi$EFT data (horizontal band) \cite{Newton:2021rni}. 
Bottom panel: Density dependence of correlation coefficient between the EoS parameters 
$K(\rho)$, $L\rho)$, $\eta(\rho)$ with crust-core transition density $\rho_t$ (top panel)}
\label{fig:RPCC}
\end{figure}
%%%%%%%%%%%%%%%%%%%%%%%%%%%%%%%%%%%%%%%%%%%%

In Fig. \ref{fig:RPCC}(a) we present the correlation between crust-core transition density with the corresponding pressure
for the full and PA EoSs of Table \ref{tab3}. In contrast to strongly correlated individual EoS, the correlation
involving all the EoSs is quite spread which can be fitted with a form $P_t \approx 1.01 (\rho_t/\rho_0)^{1.18}$.
While the predictions of $\rho_t$ and $P_t$ vary largely across different approaches, 
constraints exist on the symmetry energy imposed by isospin diffusion data in intermediate energy heavy-ion 
collisions which translates to limits on the neutron star crustal values of $0.040 \leq \rho_t \leq 0.065$ fm$^{-3}$ 
and $0.10 \leq P_t \leq 0.26$ MeV fm$^{-3}$ \cite{Xu:2009vi}. 
These limits are found to be much smaller than the predictions in the current sets of RMF and SHF models.
In the thermodynamical approach used here, the crust-core transition pressure can be 
approximated as \cite{Lattimer:2006xb,Xu:2009vi}
\begin{align}
P_t = & \frac{K \rho_t^2}{9\rho_0} \left( \frac{\rho_t}{\rho_0} - 1 \right) 
+ \rho_t \delta \Big[ \frac{1-\delta}{2} e_{\rm sym}(\rho_t)  \notag \\
& + \delta \left( \rho \frac{de_{\rm sym}(\rho)}{d\rho} \right)_{\rho_t} \Big],
\label{eq:Pt}
\end{align}
that explicitly depends on the magnitude of symmetry energy $e_{\rm sym}$ and its slope at $\rho_t$. 
The extracted limits on $P_t$ from isospin diffusion data are also found to be significantly smaller
than the fiducial value $P_t \approx 0.65$ MeV fm$^{-3}$ commonly used in several studies \cite{Lattimer:2000nx,Lattimer:2006xb}.
From Bayesian analysis of PREX measurement \cite{PREX:2021umo} of neutron skin of $^{208}\text{Pb}$ combined 
with chiral
effective field theory prediction of pure neutron matter using the liquid drop model with Skyrme energy  density functional,
a stringent constraint was estimated \cite{Newton:2021rni} to be $P_t$ is $0.33 \pm 0.07$ MeV fm$^{-3}$. On imposing this value 
in Fig. \ref{fig:RPCC}, we obtain the corresponding transition density of $\rho_t = (0.062 \pm 0.011)$ fm$^{-3}$.

%%%%%%%%%%%%%%%%%%%%%%%%%%%%%%%%%%%%%%%%%%%%
\begin{figure}[t]
\centering
 \includegraphics[width=0.95\linewidth,angle=0]{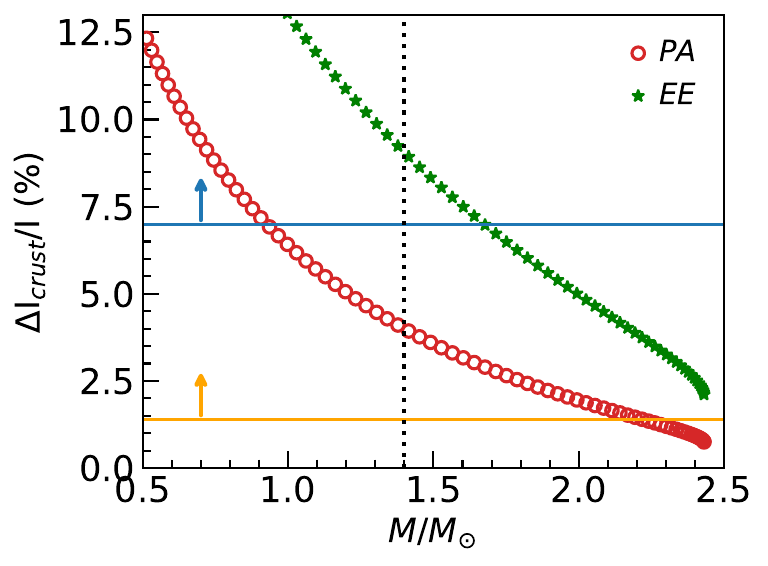}
\caption{Percentage fraction of moment of inertia of star contained in the crust 
$\Delta I_{\rm crust}/I (\%)$ as a function of neutron star mass for the BSR6 EoS \cite{Dhiman:2007ck,Agrawal:2010wg}. 
The results are in the parabolic approximation (PA) of the EoS and the exact calculation using the full EoS.}
\label{fig:CI-M}
\end{figure}
%%%%%%%%%%%%%%%%%%%%%%%%%%%%%%%%%%%%%%%%%%%%

In Fig. \ref{fig:RPCC}(b) the Pearson correlation coefficient of the transition density
with the EoS parameters $K$, $L$, and $\eta$ are depicted as a function of density calculated with the full EoSs. 
The transition density exhibits anticorrelation with the EoS parameters across the entire nuclear density range considered. 
The negative correlation, which means an increase in $L$ corresponds to a decrease in $\rho_t$, has been 
previously observed in several studies 
\cite{Xu:2009vi, Ducoin:2011fy,Alam:2015eda,Gonzalez-Boquera:2019rbh,Vinas:2021vmv, Margaritis:2021gdr}. 
The correlation between $\rho_t$ and $L$ may also be influenced by the interdependence between the EoS 
parameters $L$ and $K_{\rm sym}$ \cite{Li:2020ass} as evident from Eq. (\ref{thermo6}). 
The strong correlation between $\rho_t$ and $L(\rho_0)$ observed in Refs.
\cite{Ducoin:2010as,Alam:2015eda,Gonzalez-Boquera:2019rbh,Vinas:2021vmv} could be traced to fixed nuclear energy 
density functional used in the calculations wherein the different nuclear interactions were 
modified by varying only a few parameters of the functional. With a multitude of EOS as in the present study, 
or by either inclusion of additional term or altering the functional form, the $L-\rho_t$ correlation tends to weaken.
Although the overall crust-core transition density is round $\rho_0/2$ (see Table \ref{tab3}) which is often used as the fiducial
value in the literature, we find from Fig. \ref{fig:RPCC}(b) a moderately-large $L-\rho_t$ correlation 
of $C[\rho_t,L(\rho)] \approx 0.65$ do exist even up to densities of about $2\rho_0$.
On the other hand, a weak correlation is observed between $K$ and $\rho_t$. This can be understood from Eq. (\ref{thermo6})
where the increase of $K(\rho)$ i.e. the first two terms in the equation translates only to a small change in $\rho_t$.  As a result, the combined EoS parameter $\eta(\rho) = [K(\rho) L^2(\rho)]^{1/3}$ depicts a similar correlation as $L(\rho)$, especially at $\rho = (0.5-2)\rho_0$.

Figure \ref{fig:CI-M} presents the crustal fraction of the total moment of inertia as a function 
of neutron star mass for the BSR6 EoS. We find that more massive stars contain thin and lighter crusts 
and thereby retain smaller fraction of moment of inertia.
The parabolic approximation severely underpredicts the contribution to the crustal moment of inertia
especially for low-mass stars indicating the importance of higher order symmetry energy terms for accurate 
description of crustal properties.  
It may be mentioned that the observed frequent occurrence of abrupt spin-up episodes or glitches in Vela 
pulsar PSR B0833-45 has been explained within the vortices-pinning models by considering that some
fraction of the angular momentum is carried by the crust which translates to 
$\Delta I_{\rm crust}/I > 1.4\%$ \cite{Link:1999ca}. Accounting for the entrained neutrons in the crust enforces 
a larger constraint of $\Delta I_{\rm crust}/I > 7\%$ for the pulsar glitch spin up \cite{Andersson:2012iu}. 
As  evident from the figure, the former constraint is easily satisfied in both the full and PA
EoS for stars with $M \lesssim M_{\rm max} = 2.45 M_\odot$ and $M \lesssim 2.22 M_\odot$, respectively. 
Whereas, the latter $\Delta I_{\rm crust}/I > 7\%$ constraint restricts stars to extremely 
low mass $M \lesssim 0.93 M_\odot$ in the parabolic approximation as compared to $M \lesssim 1.67 M_\odot$ stars 
for the full BSR6 EoS.

%%%%%%%%%%%%%%%%%%%%%%%%%%%%%%%%%%%%%%%%%%%%
\begin{figure}[t]
\centering
\includegraphics[width=0.8\linewidth,angle=0]{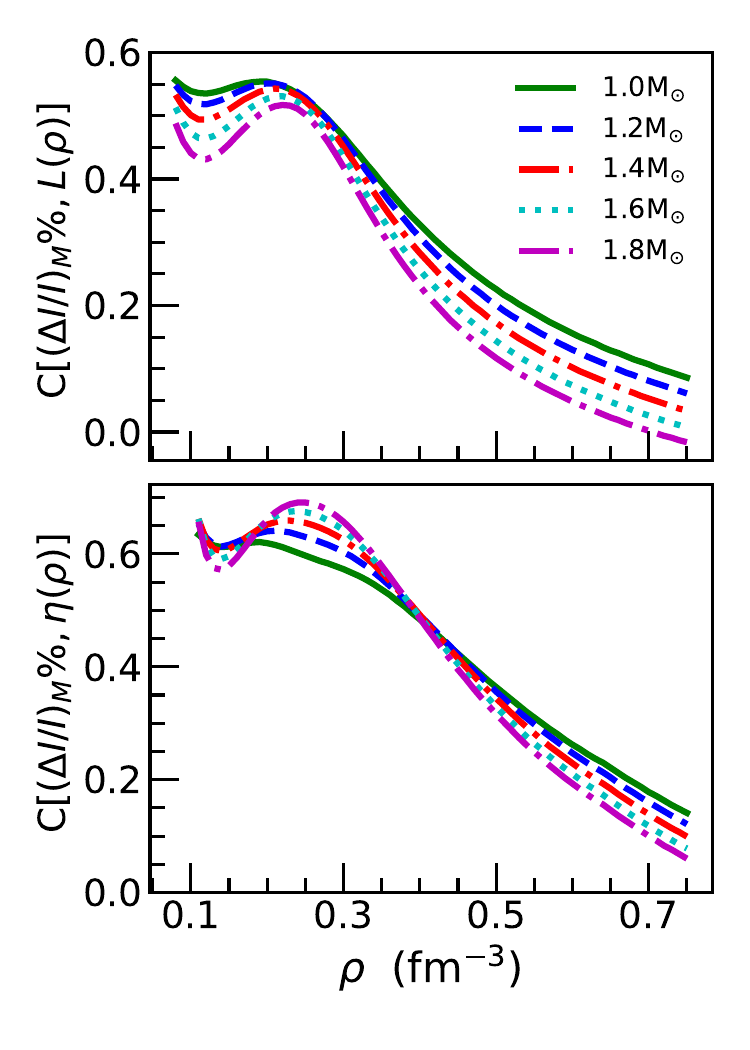}
\caption{Density dependence of correlation coefficient at fixed NS mass $M$ 
between the crustal fraction of moment of inertia of $(\Delta I_\mathrm{crust}/I)_{1.4}$ and the 
EoS parameter $L(\rho)$ (top panel) and  $\eta(\rho)$ (bottom panel) in the SHF and RMF models.}
\label{fig:dmi_den}
\end{figure}
%%%%%%%%%%%%%%%%%%%%%%%%%%%%%%%%%%%%%%%%%%%%

Figure \ref{fig:dmi_den} presents the correlation coefficient of the crustal fraction of the moment of 
inertia with the EoS parameters $L$ (top panel) and $\eta$ (bottom panel) as a function of density at fixed
values of neutron star mass $M$. The EoS parameters show a positive correlation with $\Delta I_\mathrm{crust}/I$.
The density dependence behavior with $L(\rho)$ is a reflection of Figs. \ref{fig:Corr-R-LKE} and \ref{fig:CI-M}, where the 
symmetry pressure, and hence the crustal moment of inertia, has a maximum contribution at $(0.5-2)\rho_0$ 
especially for low mass stars. Given that the stiffness or compressibility $K(\rho)$ has a larger contribution 
for massive stars on the crustal part at the subsaturation density than the slope of symmetry energy, 
their combined parameter $\eta$ exhibits correlation that increases with $M$. 
The peaks of almost all curves occur around a density of $\rho \sim 0.28$ fm$^{-3}$,
however, the correlation decreases drastically for $\rho < 0.28$ fm$^{-3}$.

%%%%%%%%%%%%%%%%%%%%%%%%%%%%%%%%%%%%%%%%%%%%
\begin{figure}[t]
\centering
 \includegraphics[width=\linewidth,angle=0]{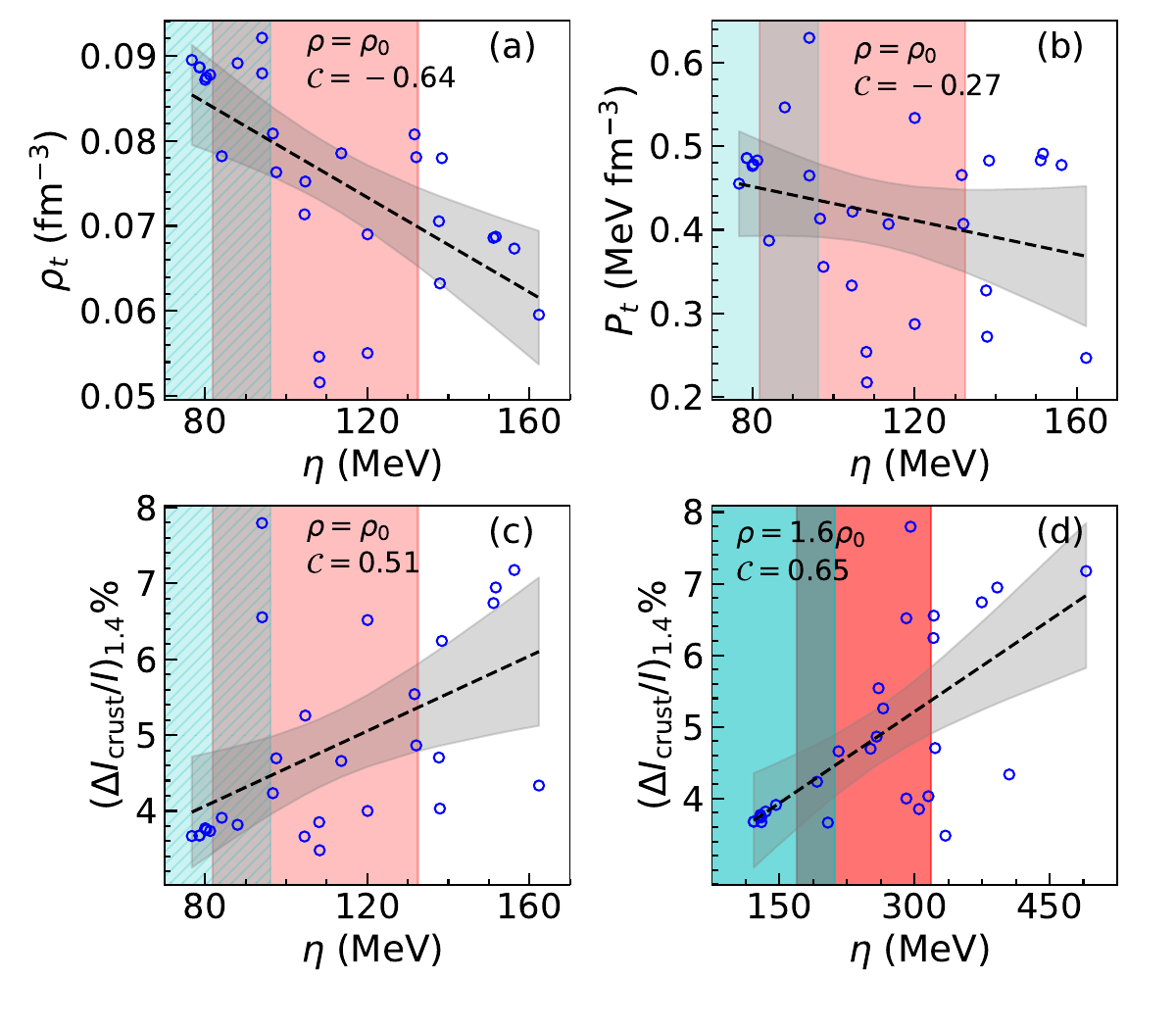}
\caption{Core-crust transition density $\rho_t$, transition pressure $P_t$, and the crustal fraction of moment 
of inertia of $1.4M_\odot$ canonical star $(\Delta I_\mathrm{crust}/I)_{1.4}$ as a function of the 
EoS parameter $\eta$ at the saturation density $\rho_0$ and $1.6\rho_0$. The bands refer to $\eta(\rho)$
bounds from Table \ref{tab2} deduced for GW170817 (blue) and GW190814 (red) events.}
\label{fig:crust_eta}
\end{figure}
%%%%%%%%%%%%%%%%%%%%%%%%%%%%%%%%%%%%%%%%%%%%

Finally, in Fig. \ref{fig:crust_eta}, the transition density $\rho_t$, pressure $P_t$ and the crustal fraction of the 
moment of inertia $\Delta I_\mathrm{crust}/I$ are presented as functions of the EoS parameter $\eta$ at 
densities $\rho_0$ and $1.6\rho_0$. While the transition pressure has a weak anticorrelation, the $\rho_t-\eta$
anticorrelation coefficient at $\rho_0$ turns out to be $-0.64$. Similarly, for the 
$\Delta I_\mathrm{crust}/I$–$\eta$, the correlation coefficients at 
$\rho_0$ and $1.6\rho_0$ are $0.51$ and $0.65$, respectively. The correlation improves for both cases at higher densities, 
with a more noticeable enhancement in the $\Delta I_\mathrm{crust}/I$–$\eta$ correlation. 
Given that the correlations between $\rho_t-\eta$ and $\Delta I_\mathrm{crust}/I$–$\eta$ are found moderate,
only approximate bound with a large uncertainty on $\rho_t$ and $\Delta I_\mathrm{crust}/I$ may be extracted by 
imposing the $\eta$ limits obtained from the $\Lambda_{1.4}-\eta$ correlation analysis
of gravitational wave events. 
To estimate the $\rho_t$ bound we have used the $\rho_t-\eta$ at $\rho_0$ of Fig. \ref{fig:crust_eta}(a), and
the crustal moment of inertia bound from $(\Delta I_\mathrm{crust}/I)_{1.4}-\eta$ at $1.6\rho_0$ of Fig. \ref{fig:crust_eta}(d),
both of which depict maximum correlation of $|\mathcal{C}| \approx 0.65$.
Hence, from Fig. \ref{fig:crust_eta} and employing the $\eta$ limits (in MeV) for GW190814, namely $81.8 \lesssim \eta(\rho_0) \lesssim 132.5$  at the saturation density provides
$\rho_t = 0.067-0.082$ fm$^{-3}$, whereas the limit $169.3 \lesssim \eta(1.6\rho_0) \lesssim 318.5$ at 
$\rho = 1.6\rho_0$ suggests the crustal moment of inertia of a 
canonical star to be $(\Delta I_\mathrm{crust}/I)_{1.4} \approx (4.1-5.4)\%$.
Similarly, for GW170817 the limits $36.2 \lesssim \eta(\rho_0) \lesssim 96.2$  
and $35.1 \lesssim \eta(1.6\rho_0) \lesssim 211.5$ MeV, give conservative estimates of $\rho_t = 0.078-0.096$ fm$^{-3}$ and $(\Delta I_\mathrm{crust}/I)_{1.4} \approx (3.0-4.5)\%$.

\section{Conclusion}
\label{sec:con}

The key parameters of the equation of state for asymmetric nuclear matter are probed at the supranuclear densities by 
examining the correlations of individual EoS parameters and their specific combinations with several 
bulk properties of the neutron stars obtained for a representative set of the nuclear energy density functionals. 
We have investigated the density dependence of these correlations to constrain the EoS parameters at densities away 
from the saturation density where the information is practically ambiguous.
The stiffness of the EoS for neutron-rich matter or neutron stars is primarily controlled by the incompressibility 
of symmetric nuclear matter $K(\rho)$ and the symmetry energy slope $L(\rho)$ which have been chosen as the basis 
for our correlation analysis. Instead of the individual parameters $K(\rho)$ and $L(\rho)$, we find their combination 
viz. $\eta(\rho) \equiv [K(\rho)L^2(\rho)]^{1/3}$ provides a stronger and reliable correlation with the neutron star radius 
$R_{1.4}$ and tidal deformability $\Lambda_{1.4}$ at suprasaturation density. Thus, $\eta$ can best describe neutron star
behavior, particularly in the high-density regime, than $K$ or $L$ independently.
We have employed (i) current simultaneous measurements of radius $R_{1.4}$ and tidal deformability $\Lambda_{1.4}$ for a 
purported neutron star of mass $1.4M_\odot$ from the secondary component of GW190814 event, (ii) the $\Lambda_{1.4}$
bound in GW170814 from the merger of binary neutron stars, and (iii) the recent $R_{1.4}$ estimate from combined 
analysis of GW170817 its electromagnetic counterparts and data from massive pulsars (GW170817+EM+PSR) to impose stringent 
bounds on the EoS parameter $\eta(\rho)$. The largest Pearson
correlation coefficient ${\cal C}[R_M,\eta] \approx {\cal C}[\Lambda_M,\eta] \approx 0.99$ was
found at the suprasaturation density of $1.6\rho_0$ which is nearly independent of the mass of neutron star.
Further, the linear combination $K + \alpha L$ and $\eta$ with neutron star radius and tidal deformability bounds
showed almost perfect correlation at the higher density $1.6\rho_0$ as compared to that at the saturation density. 
By combining the linear regression derived from both $K + \alpha L$ and $\eta$ with the neutron star properties 
($R_{1.4}$ and $\Lambda_{1.4}$), we could extract weighted averaged constraint on the compressibility 
$K(1.6\rho_0) \approx 332^{+88}_{-50}$ MeV and  
symmetry energy slope $L(1.6\rho_0) \approx 122^{+26}_{-18}$ MeV at 1.6 times the saturation density.
We also investigated the density dependence of the correlation with the crustal properties of neutron stars. 
Additionally, we examined the mass dependence of the correlation 
for the crustal fraction of the moment of inertia. The overall correlation improves when considering the 
combined EoS parameter $\eta$, while the pattern with respect to mass is primarily governed by the 
incompressibility of symmetric nuclear matter $K$.
The combined analysis of the EoS parameter $\eta$ with the crustal properties 
was found to exhibit rather moderate correlations.
This places only a conservative bound on the (GW-event weighted average) transition density of
$\rho_t = 0.076-0.087$ fm$^{-3}$ that translates to a transition pressure
of $P_t = 0.44-0.50$ MeV fm$^{-3}$, and limit the crustal moment of inertia of a
$1.4M_\odot$ neutron star to $(\Delta I_\mathrm{crust}/I)_{1.4} \approx (3.8-4.7)\%$.

\vspace{0.2cm}

\noindent \textit{\textbf{Acknowledgments:}}
The authors acknowledge financial support by the Department of Atomic Energy (Government of India) under
Project Identification No. RTI 4002.

%\bibliography{cc_ref}

\end{document}